\documentclass[5p]{elsarticle}
\usepackage{amssymb}
\usepackage{amsmath}
\usepackage{bm}
\usepackage{color}
\usepackage{import}
\usepackage{stfloats}
\usepackage{float}
\usepackage{caption}
\usepackage{xcolor}

\usepackage[overload]{empheq}
\usepackage{soul}
\colorlet{y2}{yellow!30}
\sethlcolor{y2}
\newcommand*\ybox[1]{%
	\begingroup
	\setlength{\fboxsep}{0pt}%
	\colorbox{y2}{#1}%
	\endgroup
}

\usepackage[framemethod=latex]{mdframed}

\mdfsetup{skipabove=0pt,skipbelow=0pt}
\mdfdefinestyle{ybox}{
	splittopskip=0,%
	splitbottomskip=0,%
	frametitleaboveskip=0,
	frametitlebelowskip=0,
	skipabove=0,%
	skipbelow=0,%
	leftmargin=0,%
	rightmargin=0,%
	innertopmargin=0mm,%
	innerbottommargin=0mm,%
	innerleftmargin=0mm,%
	innerrightmargin=0mm,%
	roundcorner=0mm,%
	backgroundcolor=y2,%
	hidealllines=true}
\newcommand{\MT}[2]{{\setlength{\fboxsep}{0pt}\colorbox{y2}{#2}}}

\usepackage[switch,columnwise]{lineno}

\usepackage{etoolbox} 
\makeatletter 
\newcommand*\linenomathpatch{\@ifstar{\linenomathpatch@AMS}{\linenomathpatch@}}
\newcommand*\linenomathpatch@[1]{
	\expandafter\pretocmd\csname #1\endcsname {\linenomathWithnumbers}{}{}
	\expandafter\pretocmd\csname #1*\endcsname{\linenomathWithnumbers}{}{}
	\expandafter\apptocmd\csname end#1\endcsname {\endlinenomath}{}{}
	\expandafter\apptocmd\csname end#1*\endcsname{\endlinenomath}{}{}
}
\newcommand*\linenomathpatch@AMS[1]{
	\expandafter\pretocmd\csname #1\endcsname {\linenomathWithnumbersAMS}{}{}
	\expandafter\pretocmd\csname #1*\endcsname{\linenomathWithnumbersAMS}{}{}
	\expandafter\apptocmd\csname end#1\endcsname {\endlinenomath}{}{}
	\expandafter\apptocmd\csname end#1*\endcsname{\endlinenomath}{}{}
}
\let\linenomathWithnumbersAMS\linenomathWithnumbers
\patchcmd\linenomathWithnumbersAMS{\advance\postdisplaypenalty\linenopenalty}{}{}{}
\makeatother 

\linenomathpatch{equation}
\linenomathpatch*{gather}
\linenomathpatch*{multline}
\linenomathpatch*{align}
\linenomathpatch*{alignat}
\linenomathpatch*{flalign}

\usepackage{subcaption}
\usepackage{tikz}
\usetikzlibrary{positioning}
\usepackage{hyperref}
\usepackage{cleveref}
\input{insbox}
\usepackage{cleveref}

\makeatletter
\providecommand{\doi}[1]{%
	\begingroup
	\let\bibinfo\@secondoftwo
	\urlstyle{rm}%
	\href{http://dx.doi.org/#1}{%
		doi:\discretionary{}{}{}%
		\nolinkurl{#1}%
	}%
	\endgroup
}
\makeatother

\title{Designing modular $3$D printed reinforcement of wound composite \MT{tubes}{hollow beams} with semidefinite programming}

\author[ctuMech,ctuExp]{M.~Tyburec\corref{cor1}}
\ead{marek.tyburec@fsv.cvut.cz}

\author[ctuMech]{J.~Zeman}
\ead{Jan.Zeman@cvut.cz}

\author[ctuExp]{J.~Nov\'{a}k}
\ead{novakja@fsv.cvut.cz}

\author[ctuMech]{M.~Lep\v{s}}
\ead{matej.leps@fsv.cvut.cz}

\author[ctuMech]{T.~Plach\'{y}}
\ead{plachy@fsv.cvut.cz}

\author[comp]{R.~Poul}
\ead{robin.poul@compotech.com}

\address[ctuMech]{Department of Mechanics, Faculty of Civil Engineering, Czech Technical University in Prague, Th\'{a}kurova 2077/7, Prague 6, Czech Republic}

\address[ctuExp]{Experimental Center, Faculty of Civil Engineering, Czech Technical University in Prague, Th\'{a}kurova 2077/7, Prague 6, Czech Republic}

\address[comp]{Compo Tech PLUS, s.r.o., Nov\'{a} 1316, Su\v{s}ice, Czech Republic}

\cortext[cor1]{Corresponding author}

\journal{arXiv.org}

\renewcommand\hl[1]{#1}
\renewcommand\MT[2]{{#2}}
\renewcommand\ybox[1]{%
	\begingroup
	\setlength{\fboxsep}{0pt}%
	#1%
	\endgroup}

\begin{document}

\begin{frontmatter}
\begin{abstract}
Fueled by their excellent stiffness-to-weight ratio and the availability of mature manufacturing technologies, filament wound carbon fiber reinforced polymers represent ideal materials for thin-walled laminate structures. However, their strong anisotropy reduces structural resistance to wall instabilities under shear and buckling. Increasing laminate thickness degrades weight and structural efficiencies and the application of a dense internal core is often uneconomical and labor-intensive. In this contribution, we introduce a convex linear semidefinite programming formulation for truss topology optimization to design an efficient non-uniform lattice-like internal structure. The internal structure not only reduces the effect of wall instabilities, mirrored in the increase of the fundamental free-vibration eigenfrequency, but also keeps weight low, secures manufacturability using conventional three-dimensional printers, and withstands the loads induced during the production process. We showcase a~fully-automatic \MT{pipeline}{procedure} in detail for the design, prototype manufacturing, and verification of a simply-supported composite \MT{beam}{machine tool component}, including validation with roving hammer tests. The results confirm that the 3D-printed optimized internal structure almost doubles the fundamental free-vibration eigenfrequency\hl{, allowing to increase working frequency of the machine tool}, even though the ratio between elastic properties of the carbon composite and the ABS polymer used for 3D printing exceeds two orders of magnitude.
\end{abstract}

\begin{keyword}
	Topology optimization \sep internal structure \sep semidefinite programming \sep additive manufacturing \sep elastic instabilities \sep experimental validation
\end{keyword}
\end{frontmatter}

\section{Introduction}

Offering excellent stiffness-to-weight ratios, high damping, and a~low sensitivity to fatigue and corrosion, \MT{filament wound }{}carbon fiber reinforced polymers (CFRPs) are employed in high-tech applications\hl{, including bodies of racing cars} in automotive\MT{}{~\citep{Friedrich2013}}, \hl{propulsors and turbines in} naval~\MT{}{\citep{Challis2001,Young2016}}, \hl{wing boxes and ailerons in} aerospace~\MT{}{\citep{irving2015}}, and \hl{rocket bodies in the} space industr\MT{ies}{y}~\citep{Vasiliev2012}. Considerable attention has been therefore paid to methods for optimizing \MT{their }{}structural performance\hl{ of these laminates}, particularly the laminate layup~\citep{Ghiasi2009,Ghiasi2010}. To concurrently maximize bending stiffness and keep weight low, the outer dimensions of these structures tend to be maximized, while the wall thickness is minimized. The ``thin-walledness'' of the resulting structures, combined with their anisotropy, renders them highly sensitive to shear and wall buckling instabilities manifested in low fundamental free-vibration eigenfrequencies.

Below, we review common approaches to topology optimization that reduce wall instabilities by designing an internal structure, Section~\ref{sec:to}. Section~\ref{sec:sdp} provides a brief introduction to semidefinite programming and highlights several applications in structural optimization. Finally, Section~\ref{sec:novelty} reveals the merits of designing internal structures using semidefinite programming.

\subsection{Topology optimization}\label{sec:to}
Topology optimization techniques~\citep{Bendsoe2003} provide the means for reducing wall instabilities when designing sufficiently stiff yet lightweight structures. In the simplest setting---beam cross section optimization---we search an optimal two-dimensional cross-sec\-tional shape or a~stiffening of structures whose outer shape is predefined~\citep{Kim2000}. \citet{Blasques2014}, for example, maximized the fundamental free-vibration eigenfrequency while accounting for the mass and shear center position constraints; \citet{Nguyen2018} optimized cross-sections of prismatic beams to maximize their buckling loads.

The design of optimal core sandwich structures, whose skins are stiffened by a thick core is a related challenge. For honeycomb, solid, truss, and foam rectangular panels under in-plane compression or shear loads, optimal periodic topologies can be found analytically by considering the optimality criterium of all failure modes occurring simultaneously~\citep{Vinson2005a}. For complex boundary conditions, parametric shape-opti\-mization studies are usually performed. \citet{Wang2003} studied the geometry of a~metal honeycomb sandwich beam core under torsion and bending and \citet{Xu2013} optimized the lattice core of a composite sandwich panel to increase the fundamental eigenfrequency while accounting for uncertainties in the model. They concluded that bending eigenfrequencies increase with increasing strut thicknesses, with an increase in the elastic and shear modulus of the composite, and with a~decrease in density. Although \citet{Daynes2017} optimized spatially-graded lattice structures within a single sandwich panel domain, surprisingly, almost no prior research seems to have stepped beyond parametric intuition-based designs~\citep{Birman2018,Helou2018}, the rare exception being the multi-scale topology optimization approach investigated by \citet{Coelho2015a}.

Questioning whether, where, and how to stiffen already engineered designs in order to further improve their structural performance constitutes the central question of the reinforcement problem~\citep{Olhoff1983,Diaaz1992}, superseding the former dimensional reduction and periodicity assumptions. Initial studies in this area have considered maximization of the fundamental eigenfrequency~\citep{Diaaz1992} and improving the structural frequency response of plane elastic structures~\citep{Ma1993} using the homogenization and optimality criteria methods, respectively. 

Using the ground structure approach for topology optimization of truss structures, \citet{Bendsoe1994b} fixed cross-sectional areas of a set of bars and searched for their stiffest truss reinforcement, a (non-smooth) convex quadratic programming formulation. Alternatively, the effect of a fixed boundary structure has been approximated by an appropriate application of nodal forces to the ground structure~\citep{Balabanov1996,Opgenoord2018}, but this choice influences, however, the optimized design.

In the setting of continuous topology optimization, \citet{Luo1998} developed a systematic optimization approach for the topology and orientation design of composite stiffeners of plates and shells in both static and dynamic settings, and \citet{Wang2004} optimized the overall structural rigidity of an automobile body through a maximization of the fundamental eigenfrequency. In aerospace applications, \citet{Maute2004} optimized a wing's internal structure, subjected to fluid-surface interactions; \citet{Aage2017} performed an extremely large-scale optimization of the internal structure of a Boeing 777 wing, while avoiding the traditional rib and spar designs~\citep{Stanford2014}. In military applications, topology optimization was the basis for the design of additively-manufactured lattice-reinforced penetrative warheads~\citep{Provchy2018} and for optimizing the layout weight of stiffeners in composite submarines subjected to nonsymmetric wave slap loads~\citep{Rais-Rohani2007}.

Other methods relevant to internal structure design have arisen in conjunction with recently introduced coating and infill optimization problems. \citet{Clausen2015} developed a formulation for the optimization of (uniformly) \textit{coated} structures, wherein a base material, \textit{infill}, was surrounded by another material at the interfaces, finding a~porous, complex infill significantly improves both structural buckling resistance and robustness to local perturbations when compared to optimized solid structures of equal weight and similar stiffnesses~\citep{Clausen2016,Clausen2017}. In three dimensions, optimized designs further exploit the merits of closed shell surfaces through the sandwich effect~\citep{Clausen2017}.

Inspired by natural, bone-like microstructures, \citet{Wu2017} optimized a spatially non-uniform porous infill, \citet{Wang2018} developed a sequential approach for generating graded lattice mesostructures, and \citet{Zhu2019} introduced a novel asymptotic-analysis-based homogenization approach. All these methods automatically design stiff yet porous infills for additive manufacturing products while superseding the traditional pattern-based designs~\citep{Livesu2017}. Finally,~\citet{Wu2018} extended their approach to the ultimate setting of a~concurrent optimization of coated structures and porous infills, and \citet{Groen2018} have developed a homogeni\-zation-based method to accelerate solutions.

\subsection{Semidefinite programming}\label{sec:sdp}
It has been shown in recent decades that several structural optimization problems can be modeled as se\-mi\-de\-fi\-ni\-te programs. Linear semidefinite programming (SDP) is a subset of convex optimization of the form
\begin{subequations}\label{eq:can}
\begin{align}
\min_{\mathbf{x}} \; & \mathbf{c}^\mathrm{T} \mathbf{x}\label{eq:can:obj}\\
\mathrm{s.t.}\;  	& \mathbf{X} = \mathbf{F}_0 + \sum_{i=1}^{m} x_i \mathbf{F}_i,\label{eq:affine}\\
& \mathbf{X} \succeq \mathbf{0},\label{eq:can:lmi}
\end{align}
\end{subequations}
and involves minimization of a~linear function \eqref{eq:can:obj} over a~spectrahedron, which is an intersection of an affine space \hl{{\eqref{eq:affine}}} with the cone of symmetric positive semidefinite matrices \eqref{eq:can:lmi}. In \eqref{eq:can:lmi}, the notation ``$\succeq \mathbf{0}$'' enforces positive semidefiniteness of the left hand side. Due to the linear dependence of $\mathbf{X}$ on $\mathbf{x}$ \eqref{eq:affine}, \eqref{eq:can:lmi} is commonly referred to as a~linear matrix inequality (LMI).

Applications of semidefinite programming to structural design were pioneered by~\citet{Ben-Tal2000SD}, \citet{DeKlerk1995}, and \citet{Vandenberghe1996} who developed formulations for minimum-compliance and weight truss topology optimizations. The main added value of SDP lies in its ability to effectively avoid the non-differentiability of multiple eigenvalues for free-vibrations~\citep{Ohsaki1999,Achtziger2007} and buckling~\citep{Ben-Tal2000ODT,Kocvara2002}, robust optimization~\citep{Ben-Tal1997}, and bounds improvement for optimization problems in a discrete setting~\citep{Cerveira2011}. Semidefinite programming has also found applications in optimal materials design, the Free Material Optimization approach~\citep{Ben-Tal1999}, or in the limit analyses~\citep{Bisbos2007}.

\begin{figure*}
	\centering
	\begin{subfigure}{0.45\linewidth}
		\def\svgwidth{\textwidth}\footnotesize
		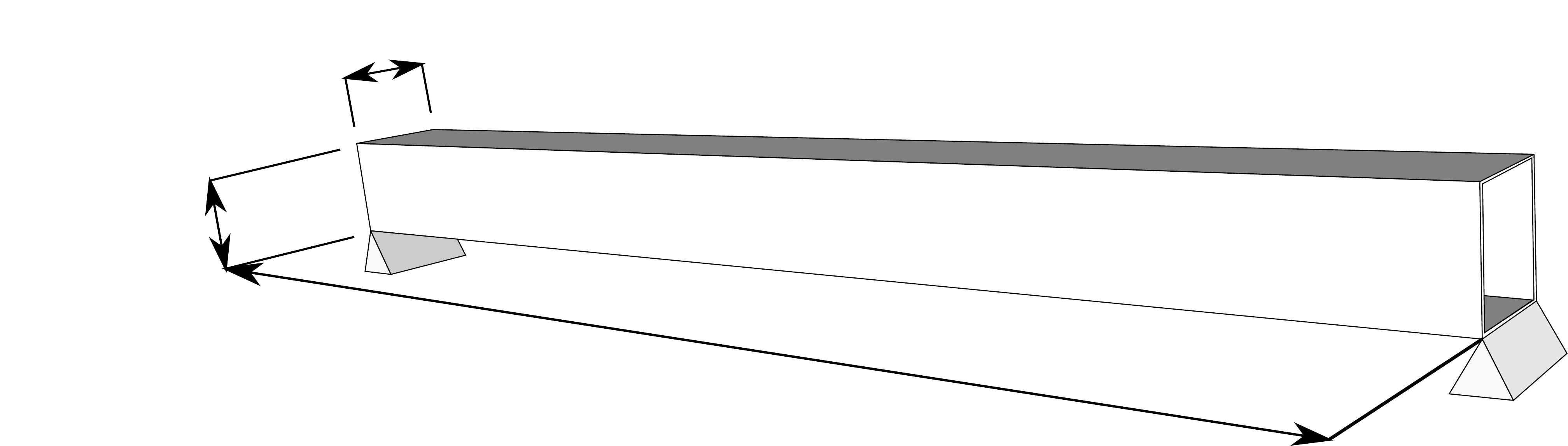
		\normalsize\caption{}
		\label{fig:dimensions_simple}
	\end{subfigure}%
	\hfill\begin{subfigure}{0.12\linewidth}
		\def\svgwidth{\textwidth}\footnotesize
\begingroup%
  \makeatletter%
  \providecommand\color[2][]{%
    \errmessage{(Inkscape) Color is used for the text in Inkscape, but the package 'color.sty' is not loaded}%
    \renewcommand\color[2][]{}%
  }%
  \providecommand\transparent[1]{%
    \errmessage{(Inkscape) Transparency is used (non-zero) for the text in Inkscape, but the package 'transparent.sty' is not loaded}%
    \renewcommand\transparent[1]{}%
  }%
  \providecommand\rotatebox[2]{#2}%
  \newcommand*\fsize{\dimexpr\f@size pt\relax}%
  \newcommand*\lineheight[1]{\fontsize{\fsize}{#1\fsize}\selectfont}%
  \ifx\svgwidth\undefined%
    \setlength{\unitlength}{322.02481842bp}%
    \ifx\svgscale\undefined%
      \relax%
    \else%
      \setlength{\unitlength}{\unitlength * \real{\svgscale}}%
    \fi%
  \else%
    \setlength{\unitlength}{\svgwidth}%
  \fi%
  \global\let\svgwidth\undefined%
  \global\let\svgscale\undefined%
  \makeatother%
  \begin{picture}(1,0.96506544)%
    \lineheight{1}%
    \setlength\tabcolsep{0pt}%
    \put(0,0){\includegraphics[width=\unitlength,page=1]{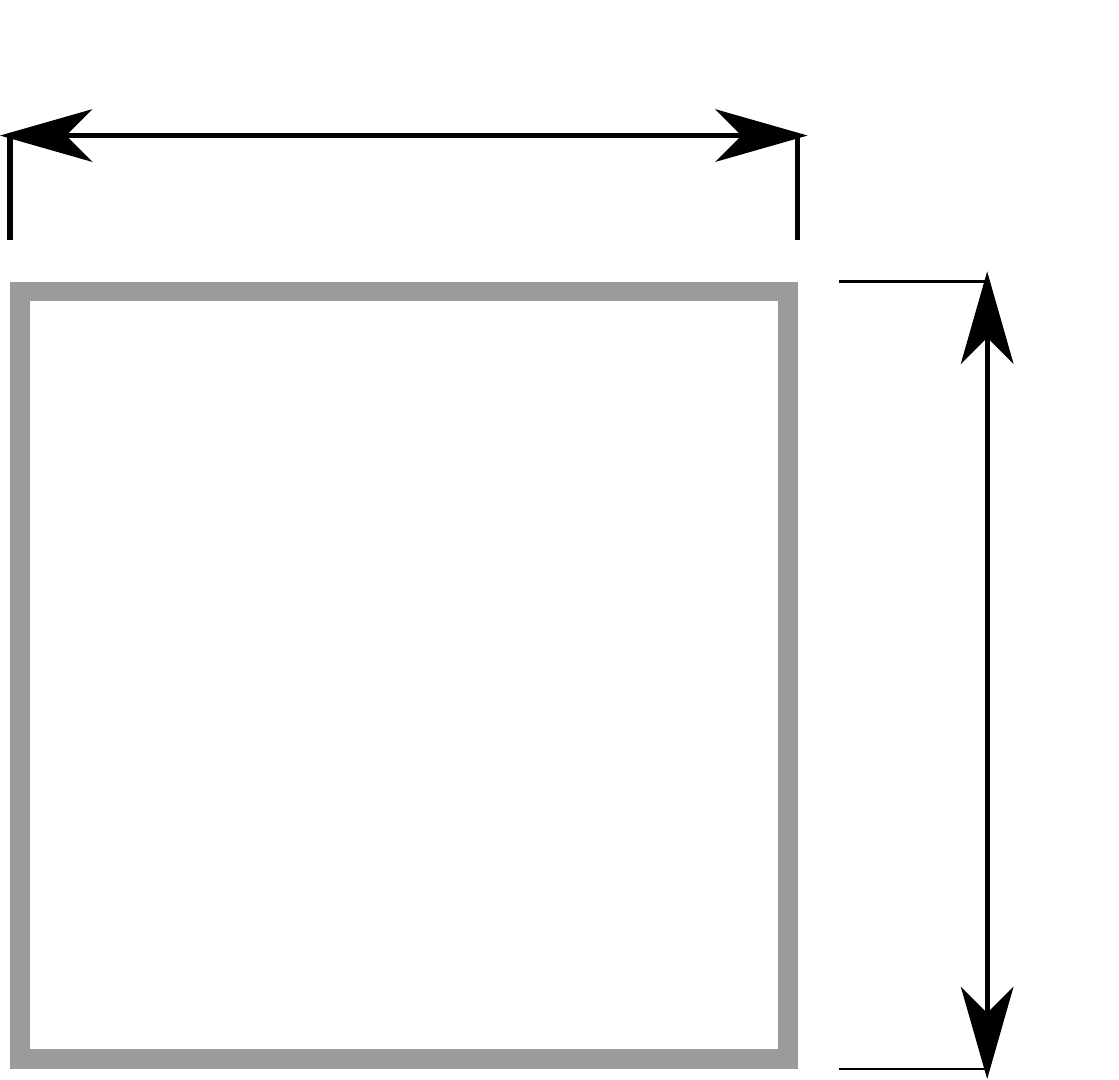}}%
    \put(0.15156611,0.86438385){\color[rgb]{0,0,0}\rotatebox{-0.29748446}{\makebox(0,0)[lt]{\lineheight{1.25}\smash{\begin{tabular}[t]{l}80 mm\end{tabular}}}}}%
    \put(0.89931835,0.59011883){\color[rgb]{0,0,0}\rotatebox{-90.29748332}{\makebox(0,0)[lt]{\lineheight{1.25}\smash{\begin{tabular}[t]{l}80 mm\end{tabular}}}}}%
    \put(0,0){\includegraphics[width=\unitlength,page=2]{dimensions2.pdf}}%
    \put(0.08771198,0.31765141){\color[rgb]{0,0,0}\makebox(0,0)[lt]{\lineheight{1.25}\smash{\begin{tabular}[t]{l}2.22 mm\end{tabular}}}}%
  \end{picture}%
\endgroup%

		\normalsize\caption{}
		\label{fig:dimensions_section}
	\end{subfigure}%
	\hfill\begin{subfigure}{0.35\linewidth}
		\def\svgwidth{\textwidth}\footnotesize
\begingroup%
  \makeatletter%
  \providecommand\color[2][]{%
    \errmessage{(Inkscape) Color is used for the text in Inkscape, but the package 'color.sty' is not loaded}%
    \renewcommand\color[2][]{}%
  }%
  \providecommand\transparent[1]{%
    \errmessage{(Inkscape) Transparency is used (non-zero) for the text in Inkscape, but the package 'transparent.sty' is not loaded}%
    \renewcommand\transparent[1]{}%
  }%
  \providecommand\rotatebox[2]{#2}%
  \newcommand*\fsize{\dimexpr\f@size pt\relax}%
  \newcommand*\lineheight[1]{\fontsize{\fsize}{#1\fsize}\selectfont}%
  \ifx\svgwidth\undefined%
    \setlength{\unitlength}{939.54273614bp}%
    \ifx\svgscale\undefined%
      \relax%
    \else%
      \setlength{\unitlength}{\unitlength * \real{\svgscale}}%
    \fi%
  \else%
    \setlength{\unitlength}{\svgwidth}%
  \fi%
  \global\let\svgwidth\undefined%
  \global\let\svgscale\undefined%
  \makeatother%
  \begin{picture}(1,0.29216375)%
    \lineheight{1}%
    \setlength\tabcolsep{0pt}%
    \put(0,0){\includegraphics[width=\unitlength,page=1]{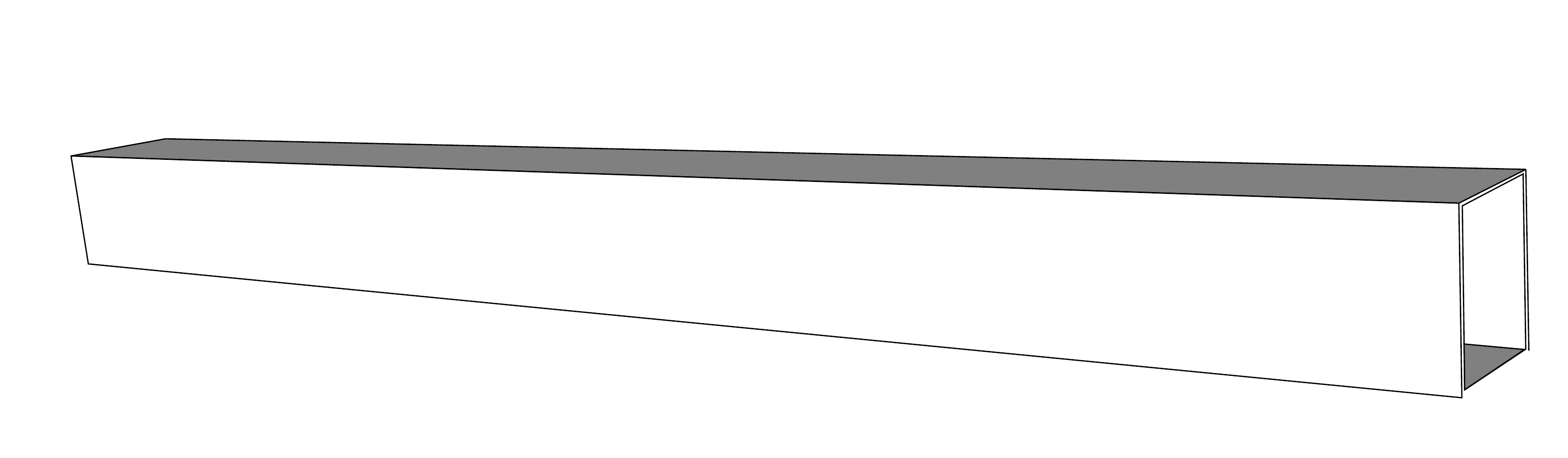}}%
    \put(0.37853833,0.24184352){\color[rgb]{0,0,0}\rotatebox{1.1343502}{\makebox(0,0)[lt]{\lineheight{59.05512238}\smash{\begin{tabular}[t]{l}$200$ kN/m$^2$\end{tabular}}}}}%
    \put(0,0){\includegraphics[width=\unitlength,page=2]{compression.pdf}}%
  \end{picture}%
\endgroup%

		\normalsize\caption{}
		\label{fig:dimensions_compression}
	\end{subfigure}
	\caption{Case study setup. (a) Outer dimensions and simply supported boundary conditions, (b) prismatic cross-section, and (c) compression molding load case.}
	\label{fig:dimensions}
\end{figure*}

\subsection{Aims and novelty}\label{sec:novelty}

In this contribution, we consider an industrial problem of designing the least-weight internal structure of a thin-walled filament-wound composite \MT{beam}{machine tool component} prone to shear and buckling wall instabilities. The beam laminate was designed for bearing dynamic loads, allowing us to describe the wall instabilities naturally in terms of free-vibrations eigenfrequencies. 

In current production process, the wall instabilities are reduced by inserting a uniform foam core structure into the beam interior, an uneconomical and labor-intensive process. Conversely, we have aimed to automatically design a structurally-efficient internal structure which can easily be manufactured using conventional low-cost $3$D printers.

To this goal, we extended the convex (linear) semidefinite programming formulation introduced by \citet{Ohsaki1999} and \citet{Ben-Tal1997} to design globally-optimal least-weight lattice-like internal structures and apply it to increasing the fundamental eigenfrequency and decreasing the compression-molding compliance of a thin-walled composite beam prototype. Note that \citet{Achtziger2007} avoided prescribed structural elements but allowed for a~non-structural mass and \citet{Ohsaki1999} did not consider prescribed mass or stiffness.

After introducing the case study of a simply-supported CFRP beam design in Section~\ref{sec:simply}, we develop its finite element representation in Section~\ref{sec:fem}. For this representation, a~semidefinite programming formulation for truss topology optimization of internal structures is developed in Section~\ref{sec:optprob}. Having designed the optimal internal structure, we post-process the optimization outputs and export, in a fully-automated way, the internal structure for additive manufacturing in Section~\ref{sec:post}. During manufacturing, the internal structure serves as the support for carbon fibers in the filament-winding production phase, and a prototype is created. Section~\ref{sec:results} describes verification and experimental validation of the prototype and concludes that its response agreed well with the model prediction.

\section{Case study}\label{sec:simply}

\begin{table*}[t]
	\centering
	\caption{Material properties of the wound composite beam laminae. $E_1$ and $E_2$ stand for the Young moduli in the fiber and transverse directions, respectively\MT{,}{;} $G_{12}$ denotes the shear modulus, $\nu_{12}$ and $\nu_{23}$ are Poisson's ratios\MT{;}{.} $\theta$ constitutes the angle between the $1$-direction and $x$, rotating around the beam surface normals. Finally, $\rho$ and $t$ denote the density and thickness of the plies.}
	\label{tab:material}
	\scriptsize
	\begin{tabular}{lrrrrrrrr}
		\hline
		Layer & $E_1$ [GPa] & $E_2$ [GPa] & $G_{12}$ [GPa] & $\nu_{12}$ [-] & $\nu_{23}$ [-] & $\theta$ [$\deg$] & $\rho$ [kg/m$^3$] & $t$ [mm] \\
		\hline
		$1$ & $128.2$ & $5.0$ & $3.4$ & $0.34$ & $0.35$ &  $89.3$ & $1,428$ & $0.25$ \\
		$2$ & $421.9$ & $3.7$ & $3.2$ & $0.37$ & $0.35$ &   $0.0$ & $1,680$ & $1.25$ \\
		$3$ & $130.9$ & $5.0$ & $3.4$ & $0.34$ & $0.35$ &  $26.9$ & $1,458$ & $0.18$ \\
		$4$ & $130.9$ & $5.0$ & $3.4$ & $0.34$ & $0.35$ & $-26.9$ & $1,458$ & $0.36$ \\
		$5$ & $130.9$ & $5.0$ & $3.4$ & $0.34$ & $0.35$ &  $26.9$ & $1,458$ & $0.18$ \\
		$6$ (casing) &   $2.0$ & $2.0$ & $0.7$ & $0.37$ & $0.37$ &   $0.0$ & $1,040$ & $0.80$\\
		\hline
	\end{tabular}
\end{table*}

As the basic structure, we consider a~prismatic, laminated composite beam \MT{of }{}$1,000$~mm\MT{}{ long}\MT{ in length}{}, with a~$80\times80$~mm thin-walled cross-section\MT{ of}{} $2.2$~mm \MT{in thickness}{thick}, Fig.~\ref{fig:dimensions_section}. According to \MT{the }{}current manufacturing technology, \MT{the }{}beam production consists of several steps, in which a~supporting structure made of manually processed high-density foam is wound biaxially with a~combination of ultra high modulus (UHM) and high modulus (HM) carbon fibers saturated with epoxy resin. The supporting structure prevents cross-section distortions induced by compression-molding loads \hl{as }shown in Fig.~\ref{fig:dimensions_compression}. Subsequently, the beam is cured, the supporting structure is pulled out, and the beam outer surface is finalized.

The final product is exposed primarily to loads that induce bending. For this purpose, most of the carbon fibers are aligned with the beam\hl{'s} longitudinal axis (layer $2$ in Table~\ref{tab:material}), denoted by $x$ in Fig.~\ref{fig:dimensions_simple}, whereas the remaining layers reduce the susceptibility to delamination. See Table~\ref{tab:material}, where all layers are listed by their orientations relative to the beam\hl{'s} longitudinal axis, $\theta$. This layered composition \MT{transmits reliably}{reliably transmits} the design forces to the supports, and is thus fully sufficient in this sense.

\begin{figure}[!b]
	\begin{subfigure}[c]{0.48\linewidth}
		\centering
		\begin{tikzpicture}
			\node (b) at (-0.5,0.5) {\includegraphics[width=0.94\linewidth]{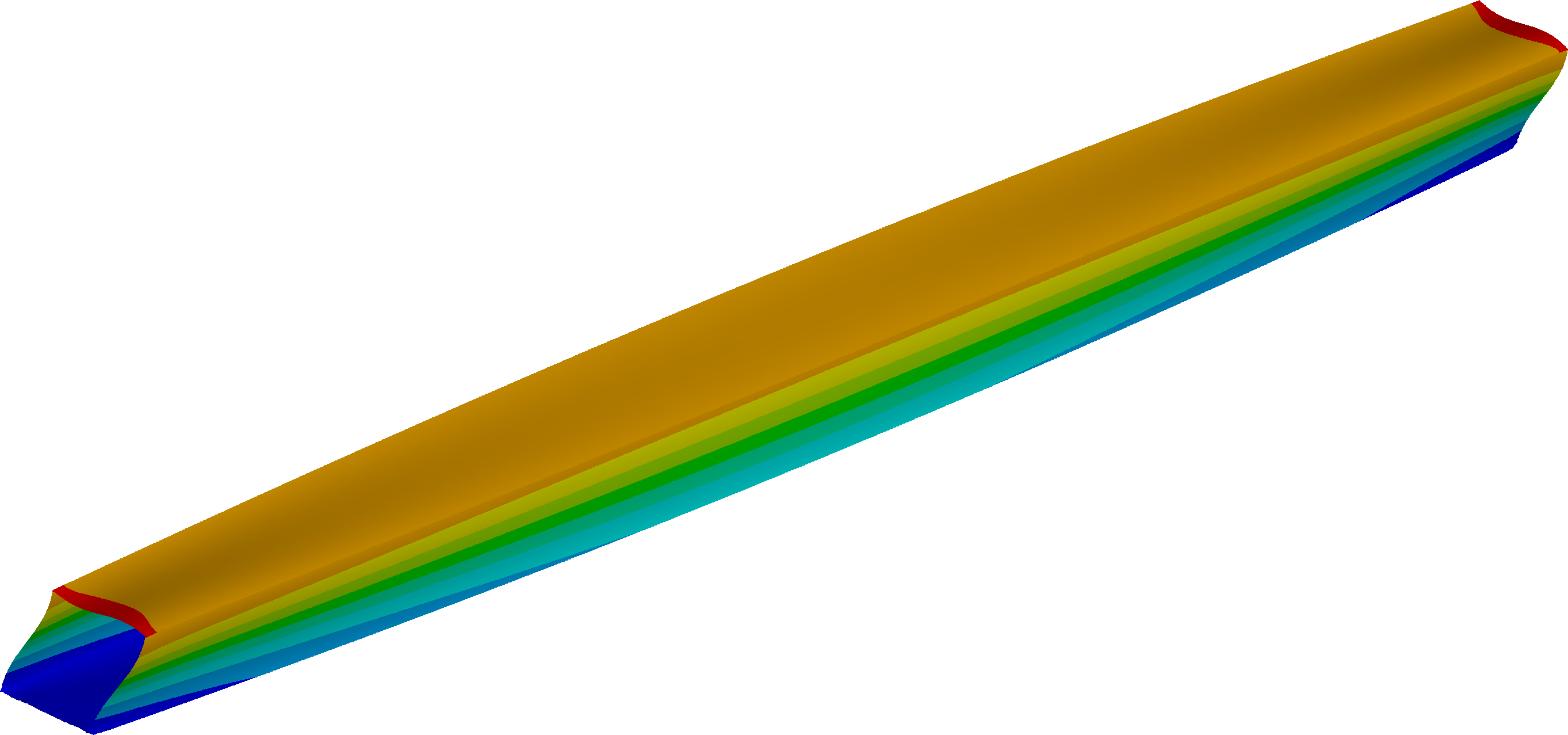}};
			\node (a) at (1.0,-0.3) {\includegraphics[width=0.18\linewidth]{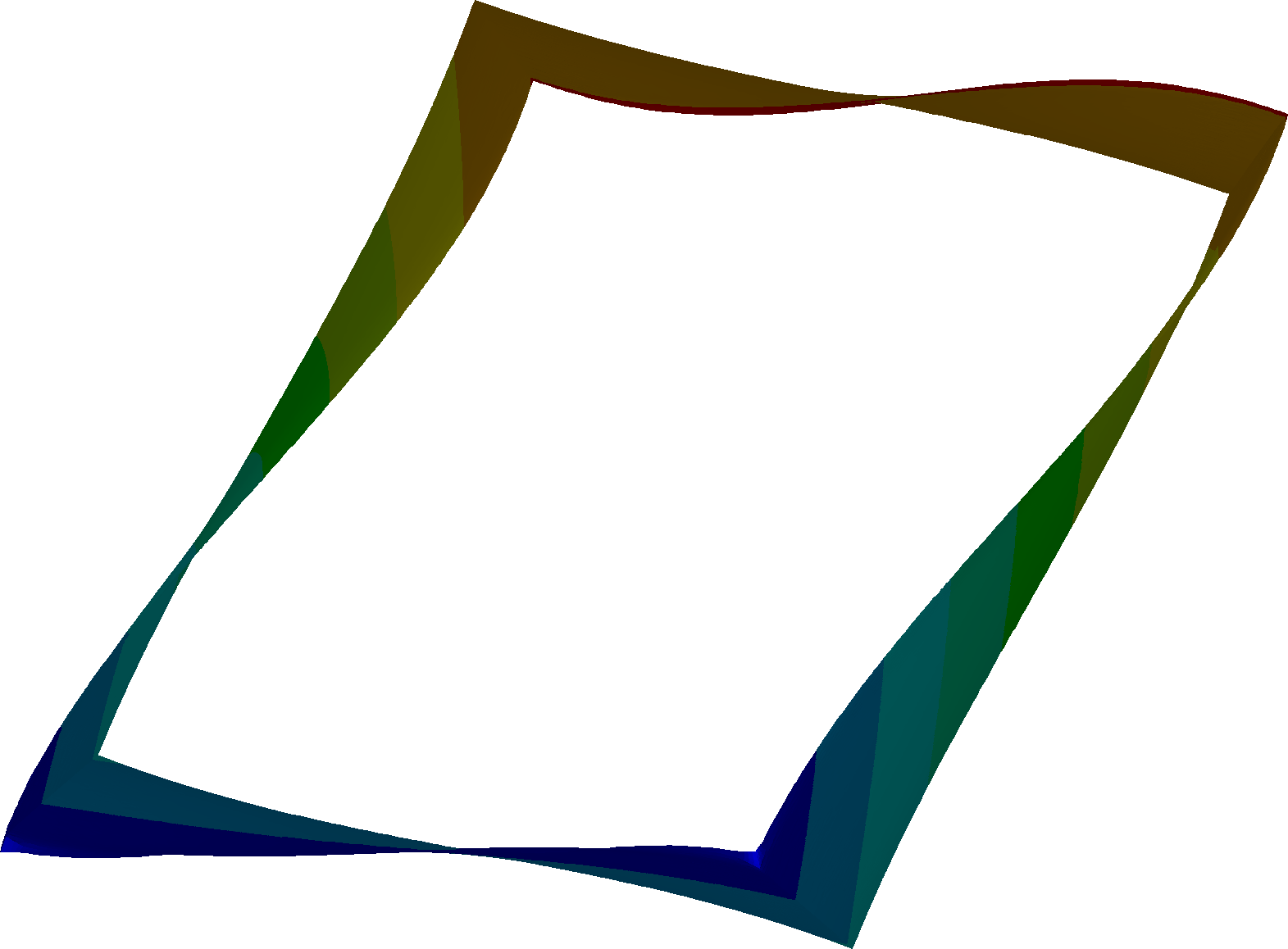}};
		\end{tikzpicture}
		\caption{First eigenmode \hl{with a frequency }of \MT{frequency }{}$128.5$~Hz.}
	\end{subfigure}%
	\hfill\begin{subfigure}[c]{0.48\linewidth}
		\centering
		\begin{tikzpicture}
			\node (b) at (-0.5,0.5) {\includegraphics[width=0.94\linewidth]{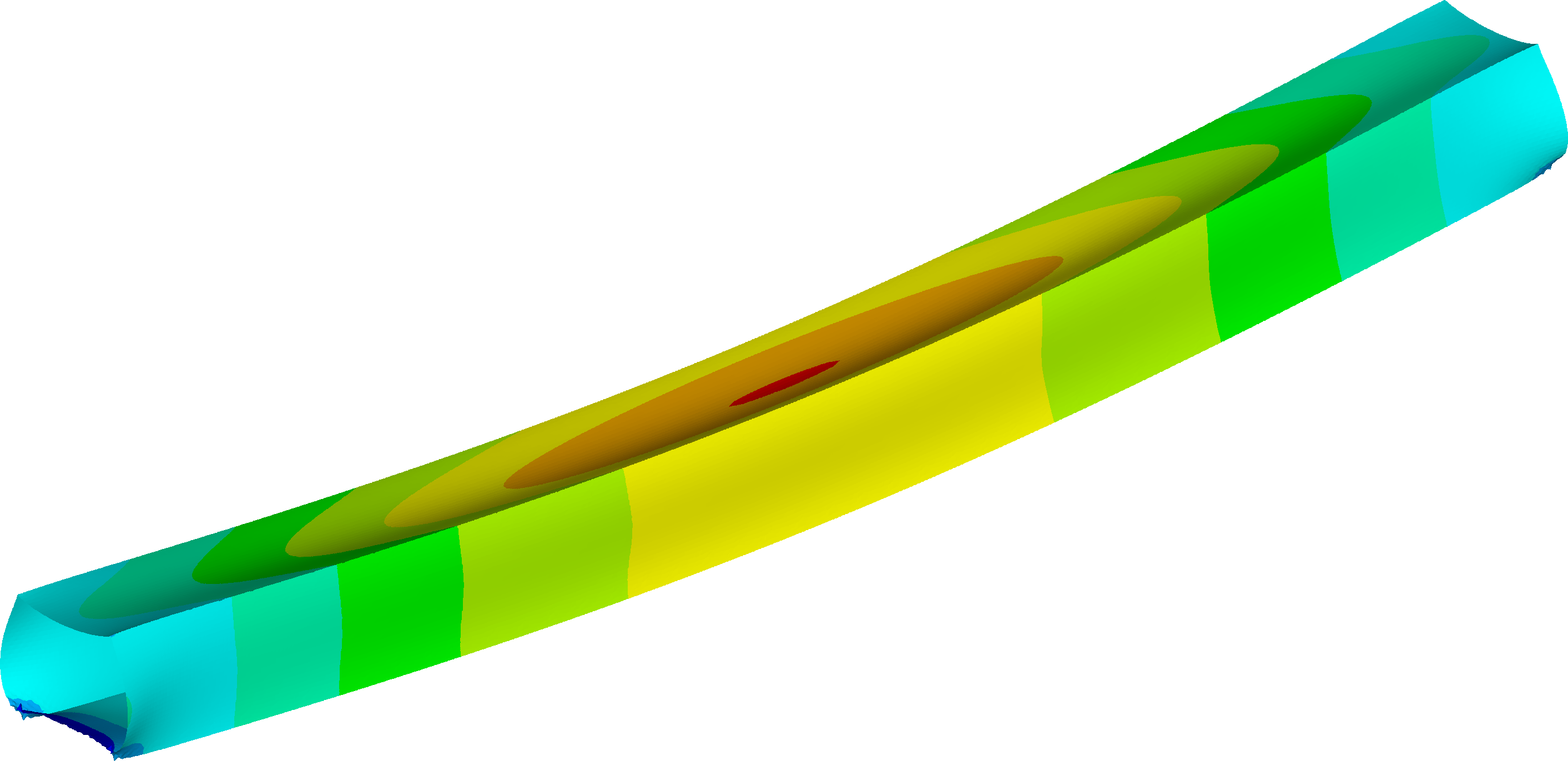}};
			\node (a) at (1.0,-0.3) {\includegraphics[width=0.18\linewidth]{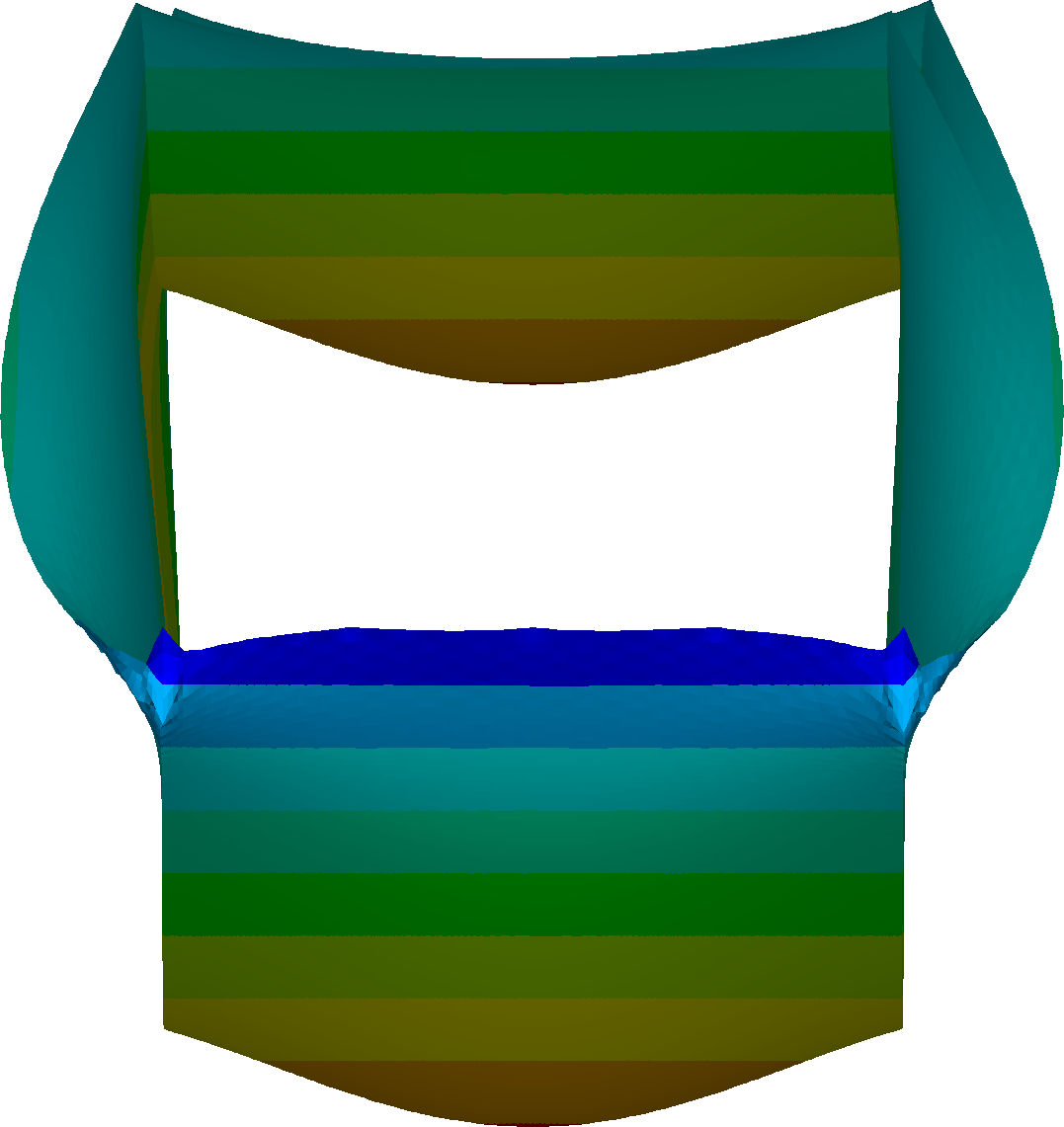}};
		\end{tikzpicture}
		\caption{Second eigenmode \hl{with a frequency }of \MT{frequency }{}$403.1$~Hz.}
	\end{subfigure}
	\caption{Axonometric and front view on the (a) first and (b) second eigenmodes of the composite beam predicted by the finite element model.}
	\label{fig:empty_simple}
\end{figure}

Attributed \MT{with}{to} transversely isotropic material properties, the beam\hl{'s} walls are, however, prone to elastic wall instabilities under shear and buckling, which also manifests in free-vibration modes and frequencies of the non-reinforced beam. Figure~\ref{fig:empty_simple} confirms that the first fundamental eigenmode\hl{ with a frequency} of \MT{frequency}{} $128.5$~Hz corresponds to shear wall instabilities, whereas the second eigenmode combines bending with buckling; all higher eigenmodes (not shown) exhibit similar wall instabilities. Because the fundamental eigenfrequency limits the maximum working frequency of the machine part, its increase is of \MT{a~}{}considerable interest.

Although the effect of these instabilities can be reduced by additional laminate layers or by \hl{also} keeping the uniform foam structure \MT{also}{ }for operational loads, the added weight, decrease in the bending eigenfrequencies, and labor-intensive production process render these approaches both time-inefficient and uneconomical.

\section{Optimal design of internal structure}\label{sec:opt}

The aim of this section is to cast the optimal internal structure design problem in the form of a linear semidefinite program \eqref{eq:can}. The internal structure has to withstand \MT{the }{}compression molding loads with a maximum deflection bound, while the internal structure is temporarily supported by a~steel mandrel passing through the beam interior, Fig.~\ref{fig:dimensions_compression}. Most importantly, the internal structure is supposed to increase the beam fundamental eigenfrequency via reduction of \MT{the }{}wall instabilities.

In this section, we first describe the finite element model of the composite beam. This finite element model serves then as the basis for establishing the optimization problem formulation, yielding an optimal internal structure design. The section is concluded by discussing\MT{ necessary}{} post-processing steps\hl{ necessary} to maintain manufacturability of the design.

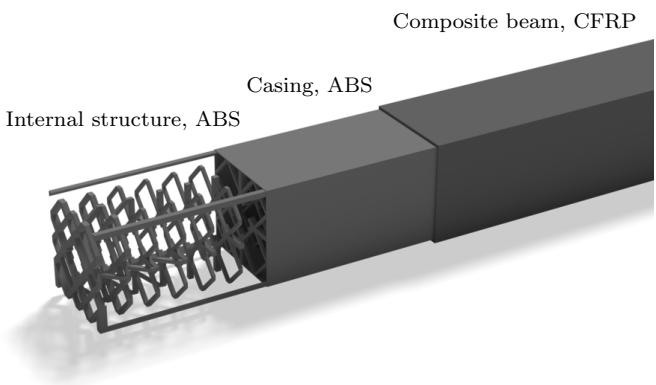
\begin{figure}[!b]
	\centering
	\begin{tikzpicture}
	\node (a) {\includegraphics[width=0.975\linewidth]{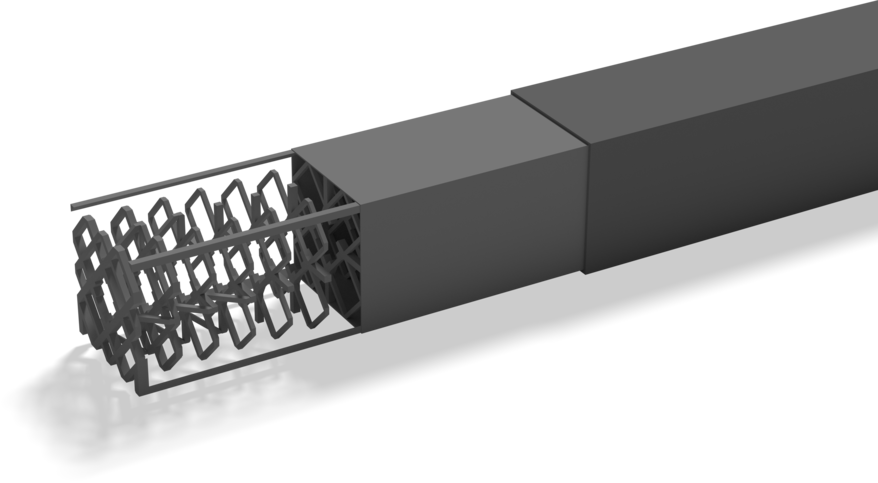}};
	\node (b) at (-2.65,1.3) {\footnotesize Internal structure, ABS};
	\node (c) at (-0.2,1.75) {\footnotesize Casing, ABS};
	\node (d) at (2.5,2.6) {\footnotesize Composite beam, CFRP};
	\end{tikzpicture}
	\caption{The entire structure of considered composite beam design: internal structure (used for the reduction of wall instabilities and for increase of the lowest free-vibration frequency); casing of the internal beam structure (to allow for wounding the final composite layer); composite layers, which transmits working load applied to the beam.}
	\label{fig:composition}
\end{figure}

\subsection{Finite element model}\label{sec:fem}

The outer composite beam surface is discretized with shell elements which are supplied with the material properties from Table~\ref{tab:material}. The beam internal structure is modeled by bar (truss) elements, with the isotropic Acrylonitrile Butadiene Styrene (ABS) material properties~\citep{Cantrell2017}: elastic modulus $E_\mathrm{ABS} = 2$ GPa, Poisson ratio $\nu_\mathrm{ABS} = 0.37$, and density $\rho_\mathrm{ABS} = 1,040$ kg/m$^3$.

\MT{A s}{S}pecial care needs to be paid to establishing a rigid connection between the internal structure and the carbon composite. The so-called \textit{casing}, see Fig.~\ref{fig:composition}, which is a~$0.8$~mm thin layer of printed beam walls, further prevents leaking \hl{of }the epoxy resin into the beam\hl{'s} interior. Casing is modeled as the bottom layer of the laminate composition, recall Table~\ref{tab:material}.

\begin{figure}[!b]
	\centering
	\def\svgwidth{4cm}\footnotesize
\begingroup%
  \makeatletter%
  \providecommand\color[2][]{%
    \errmessage{(Inkscape) Color is used for the text in Inkscape, but the package 'color.sty' is not loaded}%
    \renewcommand\color[2][]{}%
  }%
  \providecommand\transparent[1]{%
    \errmessage{(Inkscape) Transparency is used (non-zero) for the text in Inkscape, but the package 'transparent.sty' is not loaded}%
    \renewcommand\transparent[1]{}%
  }%
  \providecommand\rotatebox[2]{#2}%
  \newcommand*\fsize{\dimexpr\f@size pt\relax}%
  \newcommand*\lineheight[1]{\fontsize{\fsize}{#1\fsize}\selectfont}%
  \ifx\svgwidth\undefined%
    \setlength{\unitlength}{2041.83398438bp}%
    \ifx\svgscale\undefined%
      \relax%
    \else%
      \setlength{\unitlength}{\unitlength * \real{\svgscale}}%
    \fi%
  \else%
    \setlength{\unitlength}{\svgwidth}%
  \fi%
  \global\let\svgwidth\undefined%
  \global\let\svgscale\undefined%
  \makeatother%
  \begin{picture}(1,0.60619855)%
    \lineheight{1}%
    \setlength\tabcolsep{0pt}%
    \put(0,0){\includegraphics[width=\unitlength,page=1]{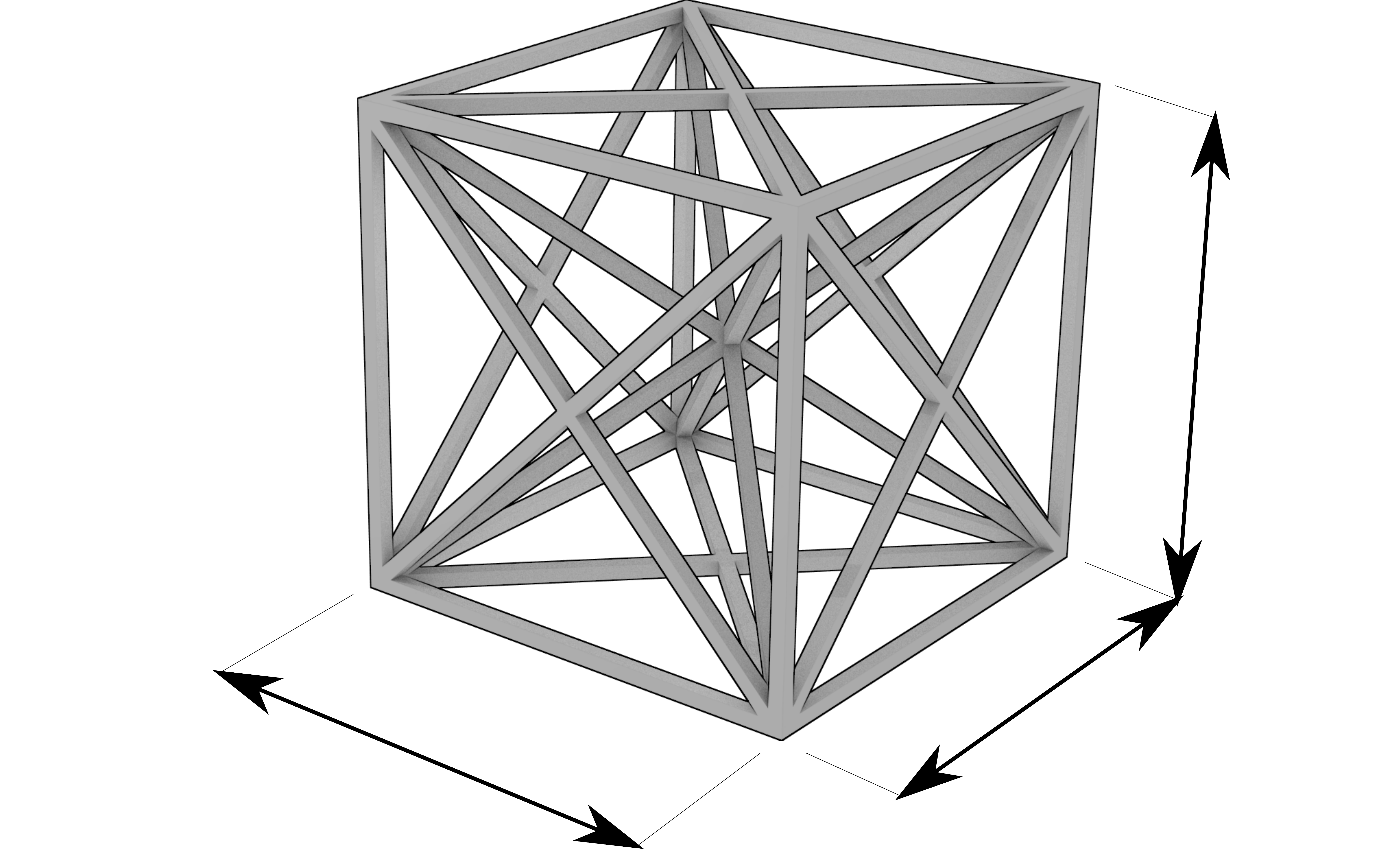}}%
    \put(0.12721712,0.05469192){\color[rgb]{0,0,0}\rotatebox{-22.689001}{\makebox(0,0)[lt]{\lineheight{1.25}\smash{\begin{tabular}[t]{l}$21.3$ mm\end{tabular}}}}}%
    \put(0.69706337,-0.04229486){\color[rgb]{0,0,0}\rotatebox{36.911021}{\makebox(0,0)[lt]{\lineheight{1.25}\smash{\begin{tabular}[t]{l}$19.5$ mm\end{tabular}}}}}%
    \put(0.86570554,0.34412383){\color[rgb]{0,0,0}\rotatebox{-3.4108303}{\makebox(0,0)[lt]{\lineheight{1.25}\smash{\begin{tabular}[t]{l}$19.5$ mm\end{tabular}}}}}%
    \put(0,0){\includegraphics[width=\unitlength,page=2]{gs.pdf}}%
    \put(0.10475146,0.17422324){\color[rgb]{0,0,0}\makebox(0,0)[lt]{\lineheight{1.25}\smash{\begin{tabular}[t]{l}$x$\end{tabular}}}}%
    \put(0.09149218,0.26714944){\color[rgb]{0,0,0}\makebox(0,0)[lt]{\lineheight{1.25}\smash{\begin{tabular}[t]{l}$y$\end{tabular}}}}%
    \put(-0.00154962,0.32068035){\color[rgb]{0,0,0}\makebox(0,0)[lt]{\lineheight{1.25}\smash{\begin{tabular}[t]{l}$z$\end{tabular}}}}%
  \end{picture}%
\endgroup%

	\caption{Ground structure building block, which fill in the entire internal volume of the composite beam to represent a (to be optimized) internal beam structure. Cross-sectional areas of individual trusses are design variables of the optimization problem~\eqref{eq:nonconvex}.}
	\label{fig:gs}
\end{figure}

The finite element model for the optimization part was developed in \textsc{Matlab}. In this model, the outer laminate was modeled with four-node \textsc{Mitc4} elements \cite{Dvorkin_Bathe_1984}. The composite beam interior was discretized into the \textit{ground structure}~\cite{Dorn1964}, a set of admissible truss\footnote{Based on comparative simulations (not shown), modeling internal structure with trusses or beams leads to \hl{an }insignificant difference in the structural response\MT{,}{} which enabled us to employ truss topology optimization approaches in Section \ref{sec:optprob}.} elements, whose cross-sections we search in the optimization part. The ground structure was constructed from $47\times4\times4$ modular building blocks shown in Fig.~\ref{fig:gs}, to guarantee manufacturability of the entire internal structure with $3$D printing. Note that the bars placed within the location of the steel mandrel were removed from the ground structure and that the shell element nodes coincided with the ground structure nodes, resulting in a rather coarse discretization of the outer layer.

\subsection{Formulation of the optimization problem}\label{sec:optprob}

\subsubsection{Non-convex formulation}

Adopting the previously described discretization, our goal is to find the cross-sectional areas $\mathbf{a}$ of $n_\mathrm{b}$ bars in the minimum-weight (or volume) ground structure, such that the fundamental eigenfrequency exceeds the user-defined lower threshold $\overline{f}$, taken as $300$~Hz in what follows, while exhibiting limit displacements $\overline{u}$ of the reinforced structure during the compression molding load case. This leads to the following optimization problem
\begin{subequations}\label{eq:nonconvex}
\begin{alignat}{4}
\negthickspace\negthinspace\min_{\mathbf{a}, \mathbf{u}_\mathrm{cm}, \mathbf{u} }\; && 
\bm{\ell}^\mathrm{T} \mathbf{a} \qquad\qquad\qquad\qquad\qquad\qquad\qquad\quad&&  && \label{eq:nonconvex:obj} \\
%
\mathrm{s.t.} \; && \negthickspace\inf_{\left( \mathbf{M}_\mathrm{fv}^\mathrm{IS}(\mathbf{a}) + \mathbf{M}_{\mathrm{fv}}^\mathrm{C} \right) \mathbf{u}\neq \mathbf{0}} \frac{\mathbf{u}^\mathrm{T} \left( \mathbf{K}^\mathrm{IS}_\mathrm{fv}(\mathbf{a}) + \mathbf{K}_{\mathrm{fv}}^\mathrm{C} \right) \mathbf{u}}{\mathbf{u}^\mathrm{T} \left( \mathbf{M}_\mathrm{fv}^\mathrm{IS}(\mathbf{a}) + \mathbf{M}_{\mathrm{fv}}^\mathrm{C} \right) \mathbf{u}} && \;\ge\; && \overline{\lambda},\label{eq:nonconvex:eigen}\\
%
&& \left(\mathbf{K}^\mathrm{IS}_\mathrm{cm}(\mathbf{a})+\mathbf{K}_{\mathrm{cm}}^\mathrm{C} \right) \mathbf{u}_\mathrm{cm} && \;=\; && \mathbf{f}_\mathrm{cm},\label{eq:nonconvex:equilibrium}\\
&& -\overline{u}\mathbf{1} \le \mathbf{u}_{\mathrm{cm,disp}} && \;\le\; && \overline{u}\mathbf{1},\label{eq:nonconvex:disp}\\
%
&& \mathbf{0} \le \mathbf{a} && \;\le\; && \overline{a}\mathbf{1},\label{eq:nonconvex:areas}
\end{alignat}
\end{subequations}
with
\begin{equation}
\overline{\lambda} = 4 \pi^2 \overline{f}^2.
\end{equation}
In this formulation, the vector $\bm{\ell}$ appearing in the objective function \eqref{eq:nonconvex:obj} collects the bar lengths in the truss ground structure. The Rayleigh quotient in \eqref{eq:nonconvex:eigen} involves\MT{ the}{} stiffness, $\mathbf{K}^\mathrm{C}_\mathrm{fv}$, and mass, $\mathbf{M}^\mathrm{C}_\mathrm{fv}$, matrices of the outer shell structure for free-vibration analysis\MT{,}{} together with\MT{ the}{} stiffness, 
$\mathbf{K}^\mathrm{IS}_\mathrm{fv}(\mathbf{a})$, and mass, $\mathbf{M}_\mathrm{fv}^\mathrm{IS}(\mathbf{a})$, matrices of the internal structure. The design-dependent contributions of the internal structure are obtained as
\begin{equation}\label{eq:fem}
	\mathbf{K}_{\mathrm{fv}}^{\mathrm{IS}}(\mathbf{a}) = \sum_{e=1}^{n_\mathrm{b}} \hat{\mathbf{K}}_{\mathrm{fv},e}^{\mathrm{IS}} a_e, \qquad 
	\mathbf{M}_{\mathrm{fv}}^{\mathrm{IS}}(\mathbf{a}) = \sum_{e=1}^{n_\mathrm{b}} \hat{\mathbf{M}}_{\mathrm{fv},e}^{\mathrm{IS}} a_e,
\end{equation}
where $\hat{\mathbf{K}}_{\mathrm{fv},e}^{\mathrm{IS}}$ and $\hat{\mathbf{M}}_{\mathrm{fv},e}^{\mathrm{IS}}$ stand for the stiffness and mass matrix of individual bars in the free-vibration (fv) setting, respectively; $a_e$ is the $e$-th component of $\mathbf{a}$ and $\overline{\lambda}$ the limit fundamental free-vibrations eigenvalue.

The constraints \eqref{eq:nonconvex:equilibrium} and \eqref{eq:nonconvex:disp} address the compression-molding (cm) load case, recall Fig.~\ref{fig:dimensions_compression}. Specifically, \eqref{eq:nonconvex:equilibrium} introduces the generalized nodal displacements $\mathbf{u}_\mathrm{cm}$ in response to the generalized load vector $\mathbf{f}_\mathrm{cm}$ corresponding to the compressive load, and $\mathbf{u}_{\mathrm{cm,disp}}$ denotes the displacement components of $\mathbf{u}_\mathrm{cm}$. The stiffness matrix corresponding to this load case consists again of the design-independent, $\mathbf{K}_\mathrm{cm}^\mathrm{C}$, and design-dependent, $\mathbf{K}_\mathrm{cm}^\mathrm{IS}(\mathbf{a})$, parts; the latter is obtained as in \eqref{eq:fem}. The symbol $\mathbf{1}$ denotes a column vector of all ones. Notice that the stiffness matrices in \eqref{eq:nonconvex:eigen} and \eqref{eq:nonconvex:equilibrium} differ because of different boundary conditions in the operational, Fig.~\ref{fig:dimensions_simple}, and manufacturing, Fig.~\ref{fig:dimensions_compression}, load cases. The constraint \eqref{eq:nonconvex:disp} requires the displacement components of $\mathbf{u}_\mathrm{cm}$ to remain smaller than the user-defined limit value $\overline{u}$, considered \MT{as}{to be} $0.5$~mm in this study. Finally, \eqref{eq:nonconvex:areas} requires the cross-sectional areas of the bars to be non-negative and smaller than $\overline{a} = 200$~mm$^2$, a~value set by the additive manufacturing constraints.

A closer comparison of the optimization problem of Eq.~\eqref{eq:nonconvex} and that of Eq.~\eqref{eq:can} reveals that the problem of Eq.~\eqref{eq:nonconvex} lacks the structure of a semidefinite program. Namely, the objective function \eqref{eq:nonconvex:obj} and the matrices in the constraints depend affinely on the design variables\hl{,} $\mathbf{a}$. However, the constraints \eqref{eq:nonconvex:eigen} and \eqref{eq:nonconvex:equilibrium} are non-convex as the stiffness and mass matrices may become singular when the zero lower-bound for cross-sectional areas is attained in \eqref{eq:nonconvex:areas}. Moreover, \eqref{eq:nonconvex:eigen} might become non-differentiable when an eigenvalue with multiplicity higher than one is encountered. Altogether, this renders the problem \eqref{eq:nonconvex} extremely difficult to solve in its original form. In the following section, we show how to re-cast the problem of Eq.~\eqref{eq:nonconvex} as a linear semidefinite programming problem.

\subsubsection{Convex semidefinite program}

\begin{figure*}[!htbp]
	\includegraphics[width=\linewidth]{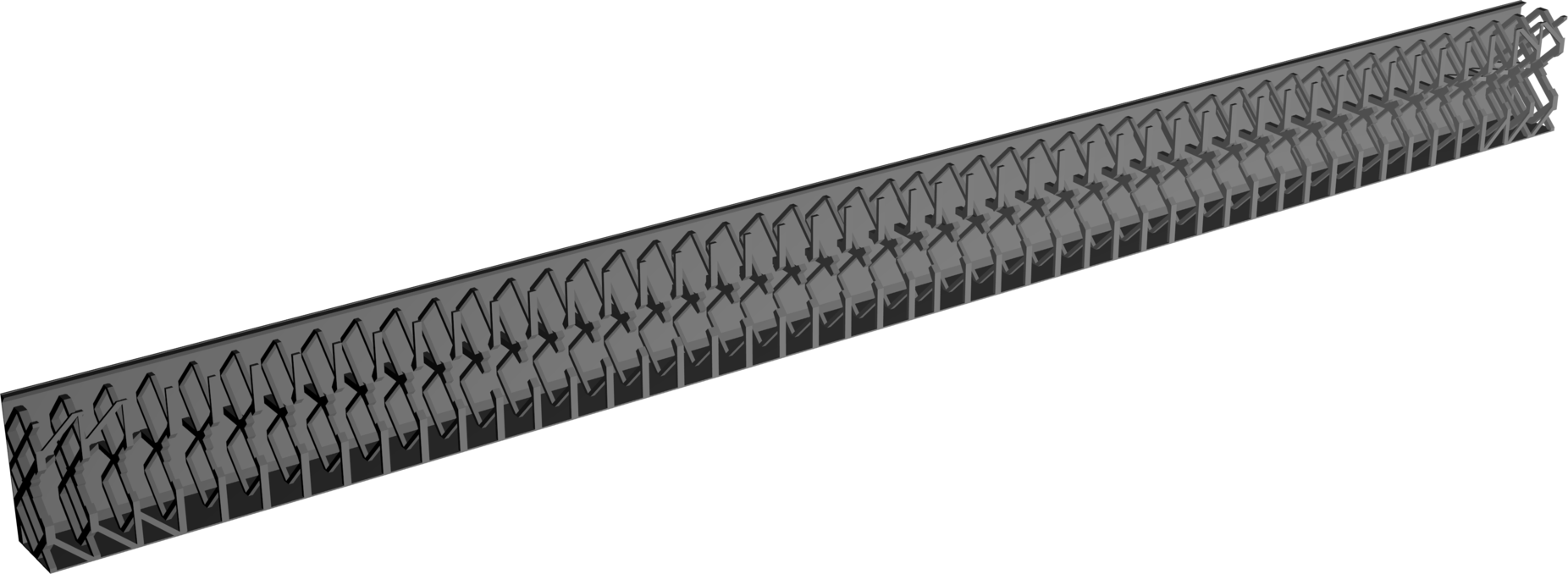}
	\caption{Symmetric half of the beam as cut off by the $xz$ plane. The top shell surface is hidden to reveal the internal structure.}
	\label{fig:cut}
\end{figure*}

Similar eigenvalue constraint\hl{s such} as \eqref{eq:nonconvex:eigen} ha\MT{s}{ve} already been studied in detail by \citet{Ohsaki1999} and \citet{Achtziger2007}. Their results allow us to rewrite \eqref{eq:nonconvex:eigen} equivalently as a~convex LMI
\begin{equation}
\mathbf{K}^\mathrm{IS}_\mathrm{fv}(\mathbf{a}) + \mathbf{K}_{\mathrm{fv}}^\mathrm{C} - 4 \pi^2 \overline{f}^2 \left(\mathbf{M}_\mathrm{fv}^\mathrm{IS}(\mathbf{a}) + \mathbf{M}_{\mathrm{fv}}^\mathrm{C}\right) \succeq \mathbf{0},
\end{equation}
where the left hand side expression is a~linear function of $\mathbf{a}$. This constraint also avoids the non-differtiability of multiple eigenvalues, see, e.g., \citep{Achtziger2007}, and effectively eliminates the kinematic variables $\mathbf{u}$ from the problem formulation.

To attain convexity of the final formulation, the compression molding constraints \eqref{eq:nonconvex:equilibrium}--\eqref{eq:nonconvex:disp} must be enforced only approximately in the form of the LMI \cite{DeKlerk1995,Vandenberghe1996,Ben-Tal1997}:
\begin{equation}
\begin{pmatrix}
c_\mathrm{cm} & -\mathbf{f}^\mathrm{T}_\mathrm{cm} \\
-\mathbf{f}_\mathrm{cm} & \mathbf{K}^\mathrm{IS}_\mathrm{cm}(\mathbf{a})+\mathbf{K}_{\mathrm{cm}}^\mathrm{C}
\end{pmatrix} \succeq \mathbf{0},
\end{equation}
in which $c_\mathrm{cm}$ denotes a prescribed upper bound on compliance (work done by external forces) of the compression molding load case. As found from parametric studies (not shown), an appropriate value of the bound is provided as
\begin{equation}
c_\mathrm{cm} = c_\mathrm{cm,0} \frac{\overline{u}}{\max\left\{ \lvert\mathbf{u}_{\mathrm{cm,disp}} \rvert\right\}},
\end{equation}
where $c_\mathrm{cm,0}$ stands for the compliance of the non-reinforced structure:
\begin{equation}
	c_\mathrm{cm,0} = \tilde{\mathbf{f}}_\mathrm{cm}^\mathrm{T} \left( \tilde{\mathbf{K}}_{\mathrm{cm}}^\mathrm{C}\right)^{-1} \tilde{\mathbf{f}}_{\mathrm{cm}}.
\end{equation}%
Here, $\tilde{\mathbf{K}}_{\mathrm{cm}}^\mathrm{C}$ and $\tilde{\mathbf{f}}_\mathrm{cm}$ are constructed from $\mathbf{K}_{\mathrm{cm}}^\mathrm{C}$ and $\mathbf{f}_\mathrm{cm}$, respectively, by application of appropriate boundary conditions. For this particular problem, this compliance bound resulted in a maximum deflection of $0.4$~mm.

The final linear semidefinite programming formulation eventually reads as 
\begin{subequations}\label{eq:sdp}
	\begin{alignat}{4}
	\min_{\MT{}{\mathbf{a}}}\; && 
	\bm{\ell}^\mathrm{T} \mathbf{a} \qquad\qquad\qquad\qquad\qquad\qquad\qquad\quad\;\;&&  && \label{eq:weight}\\
	%
	\mathrm{s.t.} \; && \mathbf{K}^\mathrm{IS}_\mathrm{fv}(\mathbf{a}) + \mathbf{K}_{\mathrm{fv}}^\mathrm{C} - 4 \pi^2 \overline{f}^2 \left(\mathbf{M}_\mathrm{fv}^\mathrm{IS}(\mathbf{a}) + \mathbf{M}_{\mathrm{fv}}^\mathrm{C}\right) && \;\succeq\; && \mathbf{0},\label{eq:eigen}\\
	%
	&& \begin{pmatrix}
	c_\mathrm{cm} & -\mathbf{f}^\mathrm{T}_\mathrm{cm} \\
	-\mathbf{f}_\mathrm{cm} & \mathbf{K}^\mathrm{IS}_\mathrm{cm}(\mathbf{a})+\mathbf{K}_{\mathrm{cm}}^\mathrm{C}
	\end{pmatrix} && \;\succeq\; && \mathbf{0},\label{eq:static}\\
	%
	&& \mathbf{1} \overline{a} \ge \mathbf{a} && \;\ge\; && \mathbf{0}.\label{eq:nonneg}
	\end{alignat}
\end{subequations}
This formulation now possesses the structure of \MT{a}{the} linear semidefinite program introduced in Section \ref{sec:sdp}, and thus can be solved efficiently via modern interior-point methods. 

For numerical solution, we adopted the state-of-the-art industrial optimizer \textsc{Mosek}~\citep{mosek}. After discretization, the problem \MT{of}{in} Eq.~\eqref{eq:sdp} has $10,216$ admissible bars in total, with the corresponding sizes of the linear matrix inequalities \MT{$7,170\times7,170$}{$5,154\times5,154$} (free-vibration, Eq.~\eqref{eq:eigen}) and \MT{$5,761\times5,761$}{$4,608\times4,608$} (compliance, Eq.~\eqref{eq:static}). \hl{After tweaking the optimization problem with the steps outlined in the following subsection, t}he optimization process itself required \MT{$41$}{$13$}~GB of memory, and terminated after \MT{$42.3$}{$5.75$} core hours running on Intel$^\text{\textregistered}$ Xeon$^\text{\textregistered}$ \MT{E5-2650v4}{Gold 6130} processors at the MetaCentrum\footnote{\url{https://metavo.metacentrum.cz/}} virtual organization cluster. The resulting distribution of the optimal internal structure is shown in Fig.~\ref{fig:cut}. Note that the internal structure increased the original weight of the beam, $1,094$~g, \MT{with}{by an} additional $488$~g ($280$~g of casing and $208$~g of reinforcing bars).

\paragraph{Improving solver \MT{convergence}{performance}} To reduce the number of iterations and time per iteration to solve \MT{the }{}problem \eqref{eq:sdp}, we rescale the cross-sectional areas to obtain the optimal values of the order of $1.0$~mm. Second, to improve both the numerical stability and convergence of the algorithm considerably, we rescale Eqs.~\eqref{eq:eigen} and \eqref{eq:static} with the square root of the Frobenius norm estimates of $\mathbf{K}_{\mathrm{fv}}^\mathrm{C}$ (Eq.~\eqref{eq:eigen}), and $\mathbf{K}_{\mathrm{cm}}^\mathrm{C}$ (Eq.~\eqref{eq:static}).
\hl{Finally, using the static condensation, Appendix~{\hyperref[app:a]{A}}, and Schur complement, Appendix~{\hyperref[app:b]{B}}, decomposition techniques, the sizes of LMIs reduce to $3,426 \times 3,426$ (free-vibration, Eq.~{\eqref{eq:eigen}}) and $2,880\times2,880$ (compliance, Eq.~{\eqref{eq:static}}). Consequently, memory usage was decreased from $21$~GB to $13$~GB, and the solution process was accelerated by $71 \%$ (from $19.5$ to $5.75$ core hours).}

\subsection{Post-processing}\label{sec:post}

Manufacturing of the optimal design is preceded by three preprocessing steps addressing individual bars, segmentation into modules, and conversion to a solid model. Note that we checked that none of the steps \MT{leads}{led} to the constraint violation and have a~rather negligible impact on the objective function, i.e., after all post-processing steps\hl{,} the internal structure volume increased from $168.3$~cm$^3$ to $175$~cm$^3$.

\begin{figure}[!b]
	\centering
	\includegraphics[width=\linewidth]{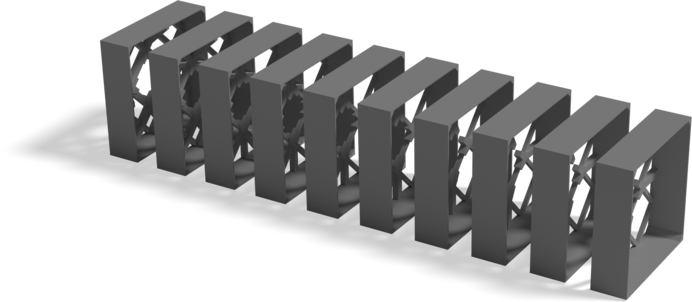}
	\caption{Segmentation of the beam internal structure.}
	\label{fig:segmentation}
\end{figure}

\paragraph{Bars post-processing}

\begin{figure*}[b]
	\centering
	\begin{tikzpicture}
	\begin{scope}
	\node (render) {\includegraphics[width=7.5cm]{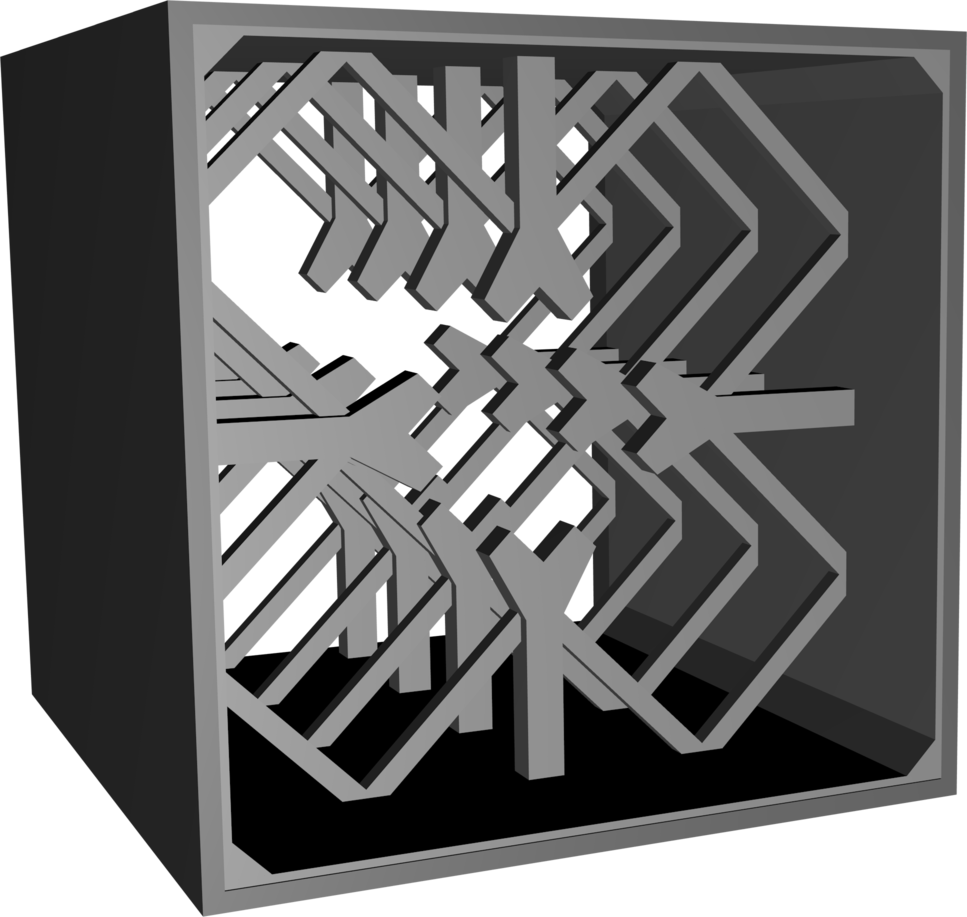}};
	\node[inner sep=0pt,below=\belowcaptionskip of render,text width=1cm,align=center]{\footnotesize(c)};
	\node[circle,color=black,thick,fill=white,inner sep=0pt,minimum size=4mm] (a2) at (-0.35,2.55) {\footnotesize a};
	\node[circle,color=black,thick,fill=white,inner sep=0pt,minimum size=4mm] (b2) at (0.47,-1.0) {\footnotesize b};
	\node[circle,color=black,thick,fill=white,inner sep=0pt,minimum size=4mm] (c2) at (1.67,2.65) {\footnotesize d};
	\node[circle,color=black,thick,fill=white,inner sep=0pt,minimum size=4mm] (d2) at (1.0,-2.1) {\footnotesize e};
	\end{scope}
	\begin{scope}[xshift=-6.7cm, yshift=2cm]
	\node (01) {\includegraphics[height=3.5cm]{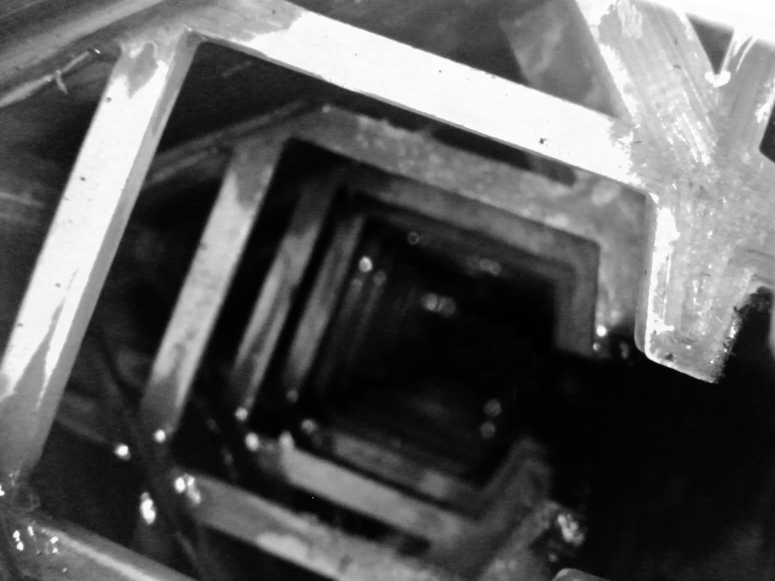}};
	\node[inner sep=0pt,below=\belowcaptionskip of 01,text width=1cm,align=center]{\footnotesize(a)};
	\end{scope}
	\begin{scope}[xshift=-6.7cm, yshift=-2cm]
	\node (02) {\includegraphics[height=3.5cm]{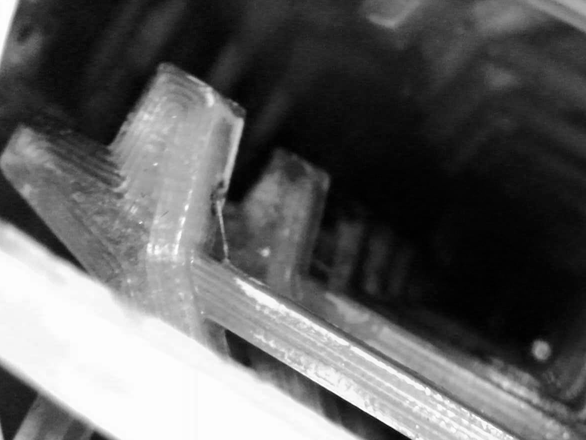}};
	\node[inner sep=0pt,below=\belowcaptionskip of 02,text width=1cm,align=center]{\footnotesize(b)};
	\end{scope}
	\begin{scope}[xshift=6.7cm, yshift=2cm]
	\node (03) {\includegraphics[height=3.5cm]{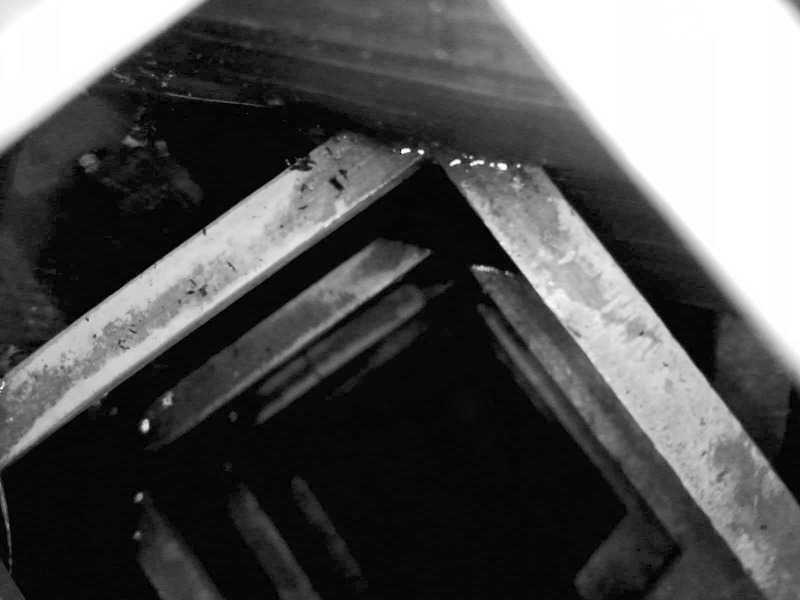}};
	\node[inner sep=0pt,below=\belowcaptionskip of 03,text width=1cm,align=center]{\footnotesize(d)};
	\end{scope}
	\begin{scope}[xshift=6.7cm, yshift=-2cm]
	\node (04) {\includegraphics[height=3.5cm]{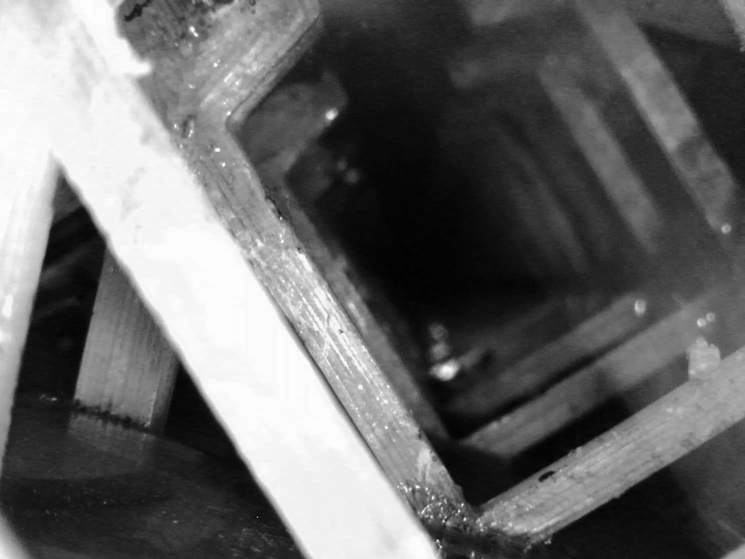}};
	\node[inner sep=0pt,below=\belowcaptionskip of 04,text width=1cm,align=center]{\footnotesize(e)};
	\end{scope}
	\end{tikzpicture}
	\caption{Endoscope camera photographs (a), (b), (d) and (e) as captured in the manufactured beam interior (c)\MT{,}{} showing that the $3$D-printed internal structure\hl{ successfully} withstood the compression-molding loads\MT{ successfully}{}.}
	\label{fig:endoscope}
\end{figure*}
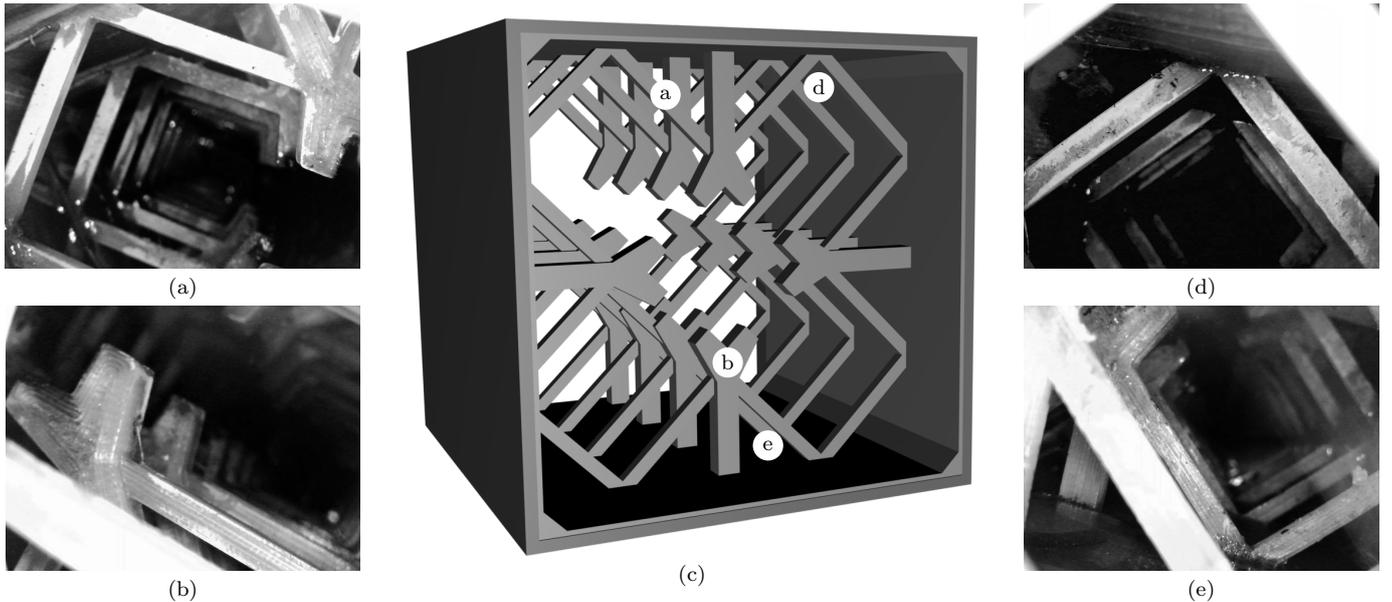

\begin{figure}[H]
	\begin{subfigure}{0.15\linewidth}
		\includegraphics[width=\linewidth]{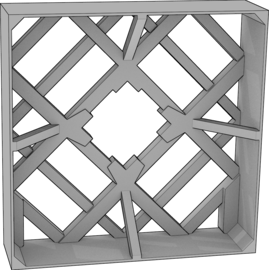}
	\end{subfigure}%
	\hfill\begin{subfigure}{0.15\linewidth}
		\includegraphics[width=\linewidth]{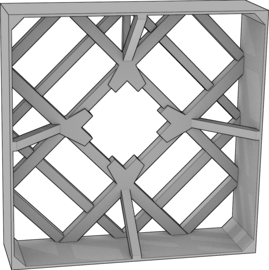}
	\end{subfigure}%
	\hfill\begin{subfigure}{0.15\linewidth}
		\includegraphics[width=\linewidth]{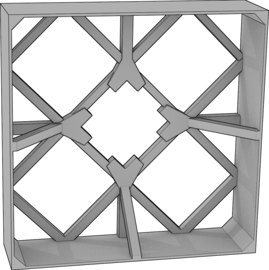}
	\end{subfigure}%
	\hfill\begin{subfigure}{0.15\linewidth}
		\includegraphics[width=\linewidth]{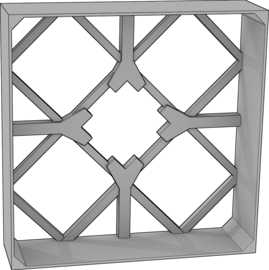}
	\end{subfigure}%
	\hfill\begin{subfigure}{0.15\linewidth}
		\includegraphics[width=\linewidth]{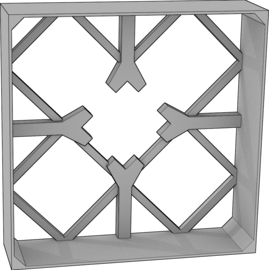}
	\end{subfigure}%
	\hfill\begin{subfigure}{0.15\linewidth}
		\includegraphics[width=\linewidth]{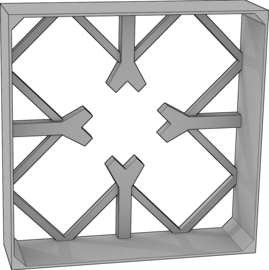}
	\end{subfigure}%
	\caption{Solid models of typical topologies of segments.}
	\label{fig:segment_top}
\end{figure}

In the initial step of \MT{the }{}module post-processing, we assign square cross-sections to each bar\MT{,}{} with the square side length according to the optimal area, $d_e = \sqrt{a_e}$. Next, we check potential intersection of bars\MT{,}{} and place a node at each intersection, which subdivides them into two and defines the new element lengths. Third, for each bar, we set the cross-sectional size $d_e$ to at least $l_e/40$, because more slender bars are difficult to manufacture with the Prusa $3$D printers used in this study. In addition to the optimized bars, the internal structure is extended with short L-shaped beams that ensure \MT{the }{}mechanical interaction between the internal structure and the steel mandrel, thus defining an empty $20.05\times20.05$~mm space along the beam longitudinal axis $x$ for its insertion, see Figs.~\ref{fig:segment_top} and~\ref{fig:endoscope}a.

\begin{figure}[!t]
	\begin{subfigure}{\linewidth}
		\hfill\includegraphics[width=2.0cm]{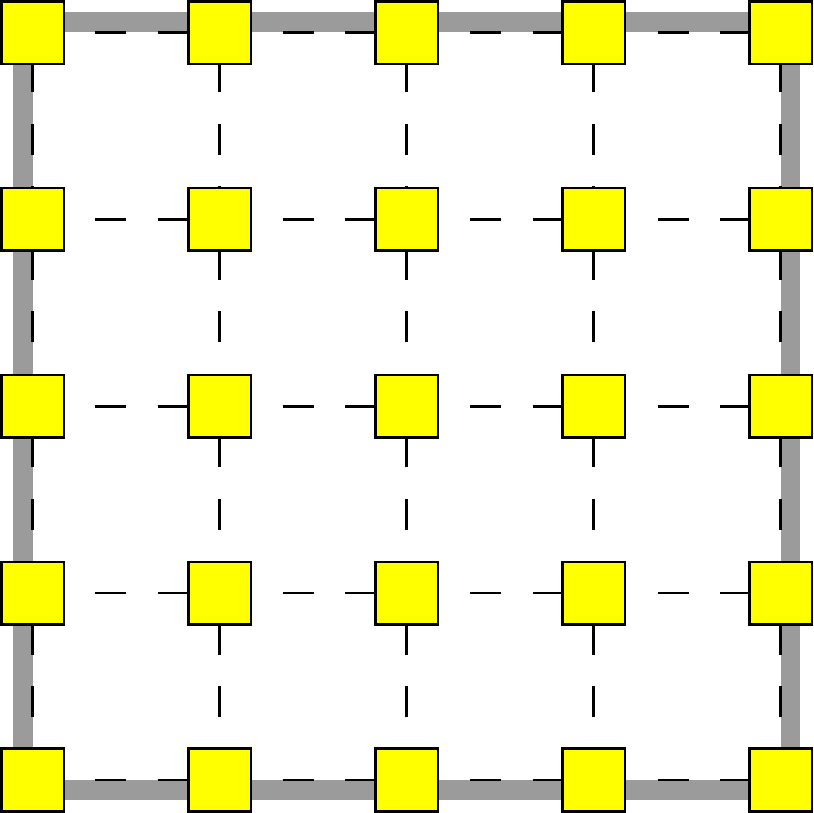}\hfill\raisebox{0.9cm}{$\overset{\scriptsize\text{re-alignment}}{\longrightarrow}$}\hfill\includegraphics[width=2.0cm]{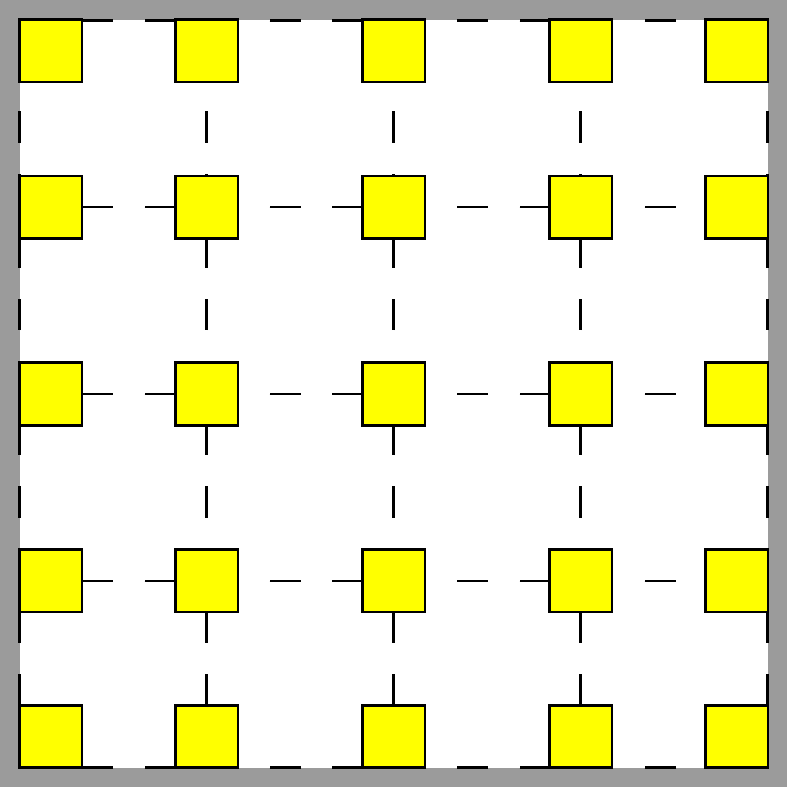}\hfill\hfill
		\caption{}
	\end{subfigure}
	\par\smallskip
	\begin{subfigure}{\linewidth}
		{\hfill\scriptsize re-alignment\hfill}
		\par\vspace*{-2mm}
		\hfill\includegraphics[width=4cm]{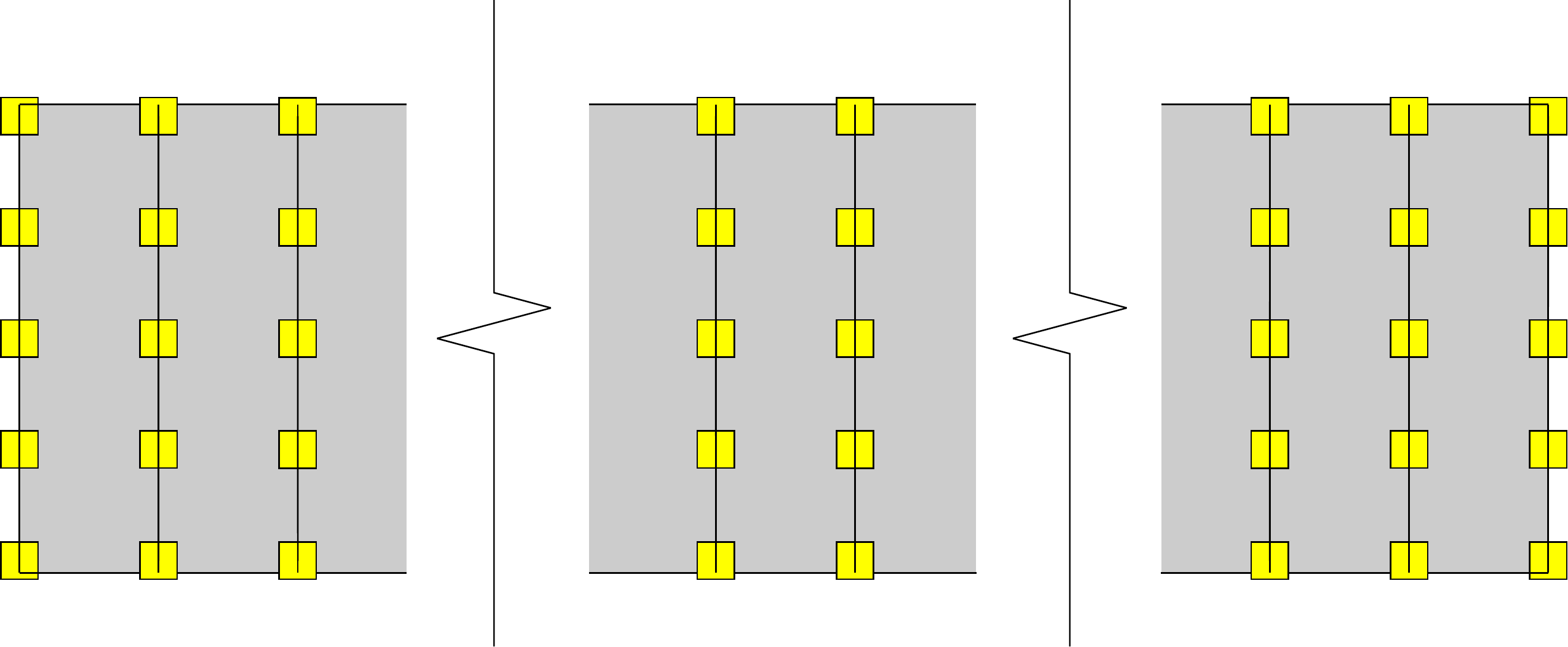}\hfill\raisebox{0.7cm}{$\longrightarrow$}\hfill\includegraphics[width=4cm]{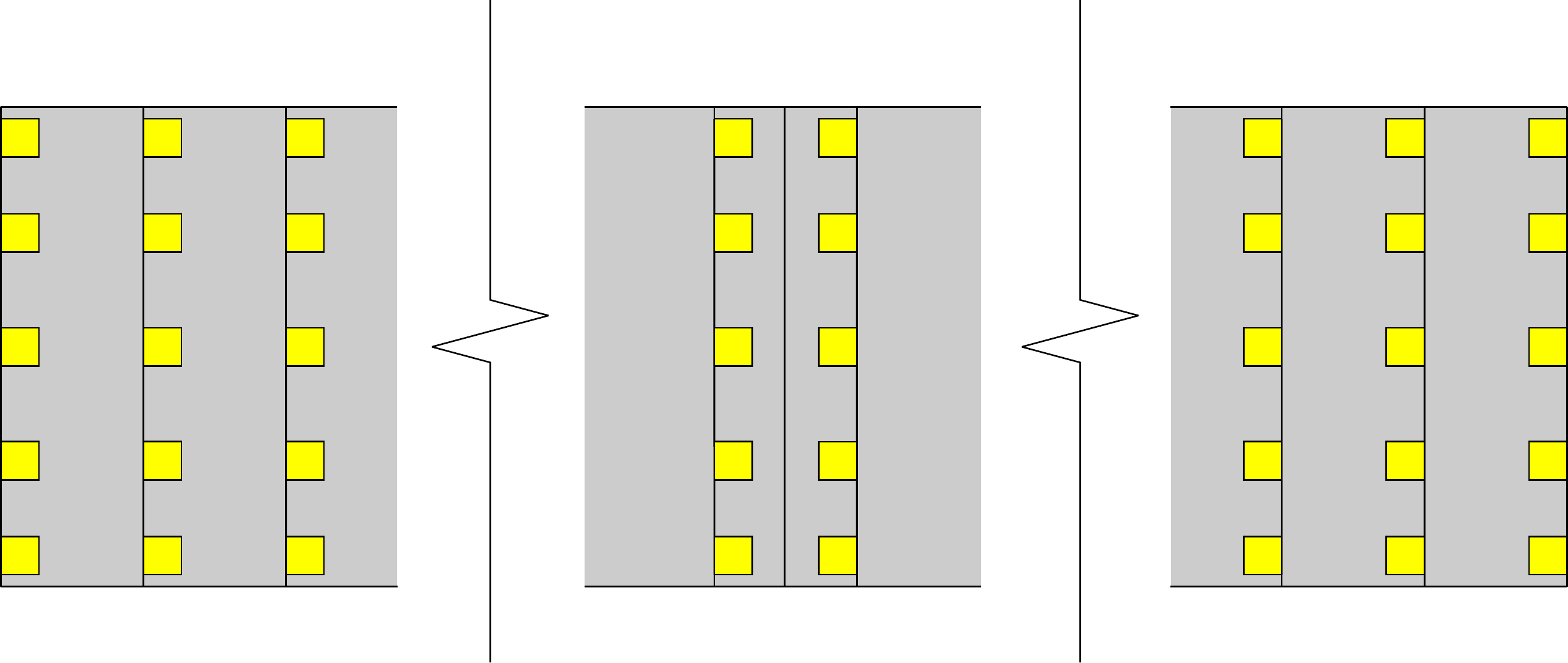}\hfill\hfill\\
		\scriptsize \hspace*{0.5mm} left end \hspace{5mm} center \hspace{4mm} right end \hspace{5mm} left end \hspace{5mm} center \hspace{4mm} right end
		\caption{}
	\end{subfigure}
	\caption{Illustration of bar cross-sections re-alignment along the beam (a) $yz$ section, and (b) longitudinal axis $x$.}
	\label{fig:alignment}
\end{figure}

\paragraph{Segmentation}

To enable parallel manufacturing with conventional 3D printers, we split the optimized internal structure into $48$ segments of approximately 20 mm in length, see Fig.~\ref{fig:segmentation} where ten selected segments are shown, to be \MT{later assembled}{assembled later} on the steel mandrel. Such segmentation requires re-alignment of bars within each beam cross-section and along the beam longitudinal axis\MT{,}{} to ensure the correct external beam dimension and a clearly defined interface among adjacent modules, see Fig.~\ref{fig:alignment} for an illustration. Note that \MT{the }{}segment production does not require any supporting material when printed along the beam longitudinal axis $x$, which would be impossible when printing the internal structure as a single-piece product.

{
\paragraph{Solid conversion} 
\MT{The a}{A}xial model conversion is performed independently and in parallel for each node of the ground\parfillskip=0pt\par}
\noindent structure. We determine first all bars attached to the considered node, elongate them by one half of their cross-sectional side lengths at both of their ends, and cut the more distant half of each of these bars off. These half-bars are then modeled by a mesh-based representation. Geometries of individual nodes then result from the mesh-boolean operations\MT{,}{} performed with the \textsc{Cork}\footnote{\url{https://github.com/gilbo/cork}} library. Finally, the overall segment geometry consists of the union of all nodal geometries, see Fig.~\ref{fig:segment_top} for typical topologies of post-processed segments, and can be readily exported to\MT{ the}{} patch-based \textsc{Stl} file format, for example.

\section{Results}\label{sec:results}

\subsection{Manufacturing}

After the automated export of the optimized internal structure into\MT{ the}{} \textsc{Stl} format, the part was additively manufactured using the Fused Deposition Modeling method with Prusa i3 MK3 printers. Printed segments were inserted on a~$20\times20\times1,200$~mm steel mandrel of $1.5$~mm wall thickness, with its surface lubricated with \MT{v}{V}aseline to simplify the pull-out process, and connected with acetone etching and a thin layer of epoxy glue.

\begin{figure}[!t]
	\centering
	\includegraphics[width=0.9\linewidth]{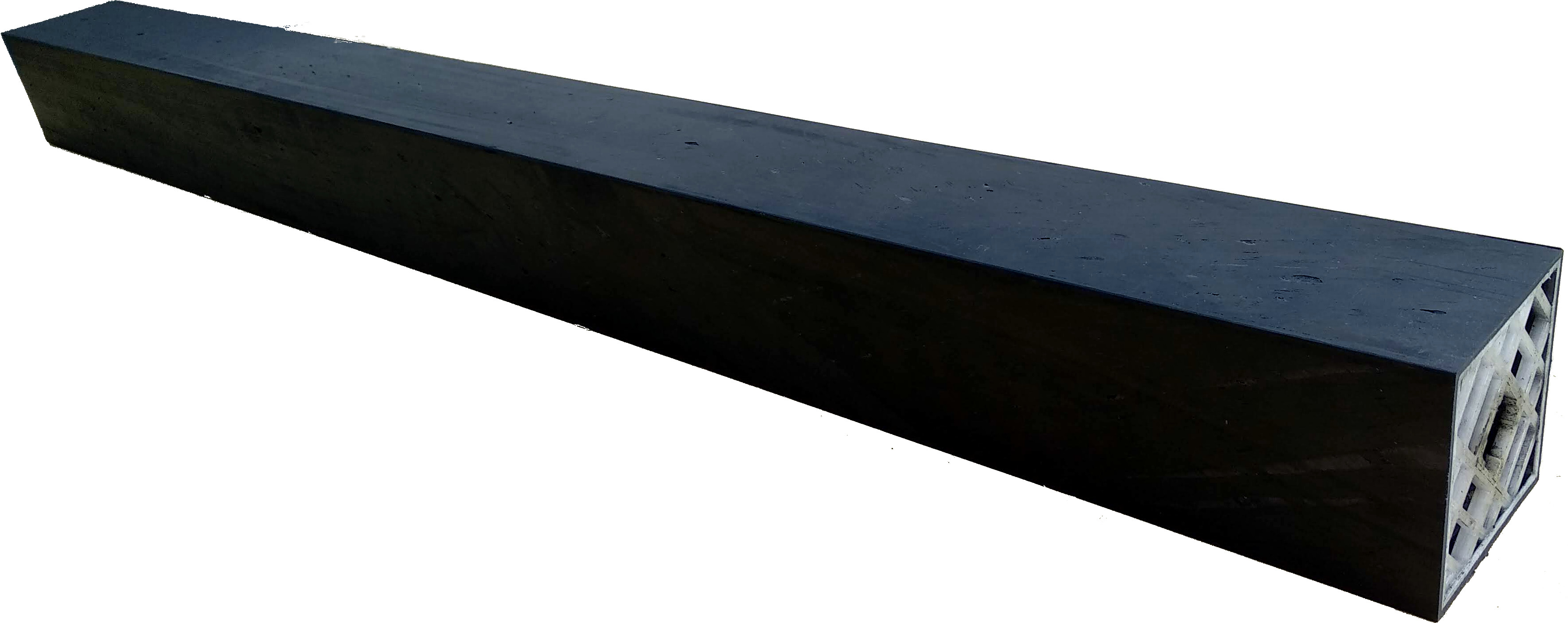}
	\caption{Manufactured prototype of a composite beam with optimized stiffening internal structure.}
	\label{fig:prototype}
\end{figure}

\hl{The prototype beam was produced by CompoTech Plus company using the filament winding technology with axial fiber placement. This technology relies on the positioning of the tows of carbon fibers impregnated by the epoxy resin on the casing, placed in specified directions and specified quantity to reach expected dimensions and mechanical properties of the final product. The casing defines the internal shape of the beam and acts as an internal mold. After the fiber placement operation, the product (with still liquid resin) is placed into the press, the outer shape is formed, and the composite is consolidated. In the press, the product hardens at the room temperature. Finally, the prototype, Fig.~{\ref{fig:prototype}}, is postcured at the elevated temperature of $90^\circ$C.} 

\MT{S}{The s}uccessful manufacturing process was followed by inspection of the prototype using an endoscope camera. Video and photograph sequences, see Fig.~\ref{fig:endoscope}, revealed that the internal structure successfully withstood the compression molding pressure without any significant visual defects. The minor deviations from the assumed model reside in a small amount of \MT{v}{V}aseline residue\MT{s}{} and \MT{a seldom}{slight} leakage of\hl{ the} epoxy resin through casing interfaces. Another difference appeared in \hl{the }increased outer dimensions of the beam, $0.32$~mm on average, caused by an insufficiently closed press cover. The total prototype weight of $1.768$~kg \MT{is}{was} therefore \MT{by}{ }$186$~g higher than the model predictions due to the additional epoxy resin.

\subsection{Verification}\label{sec:verification}

\begin{figure}[!t]
	\begin{subfigure}[c]{0.48\linewidth}
		\centering
		\begin{minipage}{\linewidth}
			\begin{tikzpicture}
			\node (b) at (-0.5,0.5) {\includegraphics[width=0.94\textwidth]{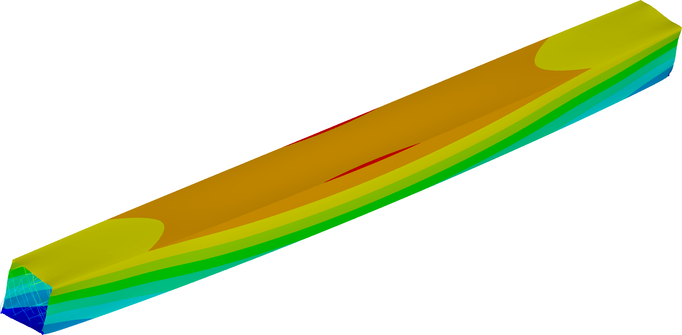}};
			\node (a) at (1.0,-0.3) {\includegraphics[width=0.18\textwidth]{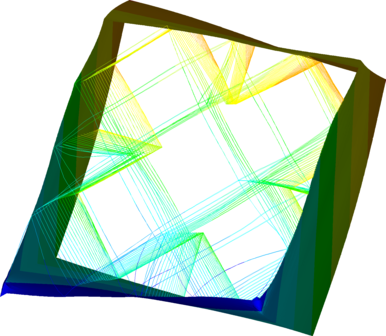}};
			\end{tikzpicture}
		\end{minipage}
		\caption{First eigenmode\hl{ with a frequency} of\MT{ frequency}{} $246.7$~Hz.}
	\end{subfigure}%
	\hfill\begin{subfigure}[c]{0.48\linewidth}
		\centering
		\begin{tikzpicture}
		\node (b) at (-0.5,0.5) {\includegraphics[width=0.94\textwidth]{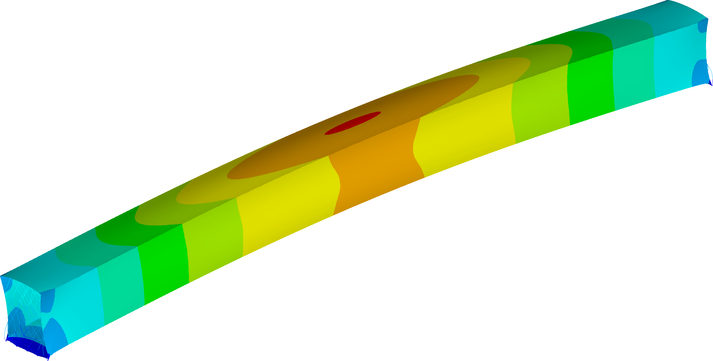}};
		\node (a) at (1.0,-0.3) {\includegraphics[width=0.15\textwidth]{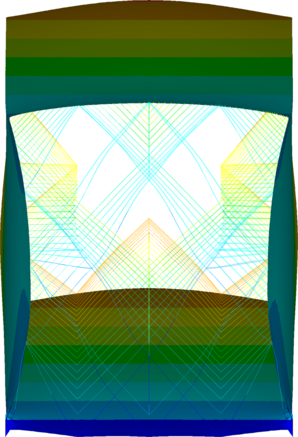}};
		\end{tikzpicture}
		\caption{Second eigenmode\hl{ with a frequency} of\MT{ frequency}{} $347.0$~Hz.}
	\end{subfigure}
	\caption{Axonometric and front view on the (a) first and (b) second eigenmodes of the reinforced composite beam predicted by the refined finite element model.}
	\label{fig:internal_simple}
\end{figure}

Recall that in the optimization we required increasing the fundamental free-vibration eigenfrequency above $300$~Hz. To check this value, we employed an independent model in \textsc{Ansys}. Compared to the model used for optimization, this model employs the dimensions measured in-situ, more refined discretization of the outer shells (element type \textsc{Shell181}), and models the internal structure with beam elements (\textsc{Beam188}) instead of trusses. Besides, the composite shells are supplemented with an additional layer of epoxy resin to account for the increased epoxy content. As a result, the model predicts that the beam fundamental eigenfrequency was increased by $92$\MT{~}{}\% from $128.5$~Hz to $246.7$~Hz, compare Figs.~\ref{fig:empty_simple}~and~\ref{fig:internal_simple}. The effect of wall instabilities was reduced jointly in all the remaining eigenmodes (not shown).

Even though the fundamental eigenfrequency did not exceed the limit value, we find these results satisfactory because of two reasons: First, we attribute this discrepancy mainly to the manufacturing imperfections, which can be attributed to the prototype character of the manufacturing process and can be easily resolved in \MT{a~}{}serial production. Second, the constraint violation is comparable to the difference between numerics and experiments as shown in the next section.

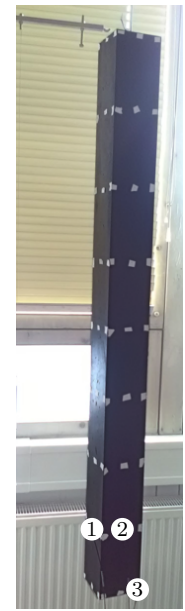
\begin{figure}[!b]
	\centering
	\begin{tikzpicture}
	\node (a) {\includegraphics[height=8cm]{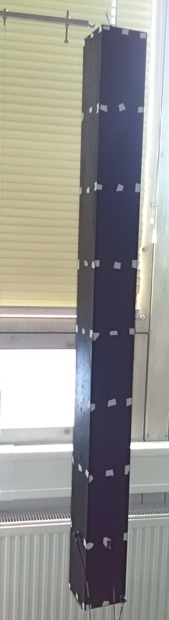}};
	\node[circle,color=black,thick,fill=white,inner sep=0pt,minimum size=3mm] (a2) at (0.5,-3.75) {\footnotesize 3};
	\node[circle,color=black,thick,fill=white,inner sep=0pt,minimum size=3mm] (a2) at (0.3,-2.95) {\footnotesize 2};
	\node[circle,color=black,thick,fill=white,inner sep=0pt,minimum size=3mm] (a2) at (-0.1,-2.95) {\footnotesize 1};
	\end{tikzpicture}
	\caption{Free-free-vibration validation setup. Locations of $54$ impact points are indicated by gray squares and positions of $3$ accelerometers are marked by white circles.}
	\label{fig:hammer}
\end{figure}

\subsection{Validation}

Dynamic response was validated with the roving hammer test in the free-free-vibration setting because it eliminates the need to reproduce the simply supported kinematic boundary condition in the experiment. To this goal, the beam was suspended at one of its ends, three piezoelectric acceleration transducers Type 4507B005 Br\"{u}el\&Kjaer were placed on the beam\hl{'s} outer surface, two of which were located in the middle of adjacent sides of the beam\hl{'s} cross-section \MT{in}{at} one\MT{ }{-}eighth of the beam\hl{'s} length, and the third one was placed \MT{in the beam}{at the} corner\hl{ of the beam}, Fig.~\ref{fig:hammer}. Two adjacent sides of the beam surface were marked with a~regularly spaced grid of $54$ points, $27$ on each side. These points then served as the excitation points for the impact hammer Type 8206 Br\"{u}el\&Kjaer equipped with \MT{the}{a~}force transducer.

\MT{The m}{M}easurement was realized \MT{by}{using} data acquisition front-end hardware Type 3560B Br\"{u}el\&Kjaer. The frequency response functions (FRFs) were evaluated from the recorded response (acceleration) and excitation (force) using the Fast Fourier Transform for all 54 points. The natural frequencies and mode shapes were evaluated from the FRFs with\MT{ the}{} MEscope software developed by\hl{ the} Vibrant Technology company.

\begin{figure*}[!htbp]
	\centering
	\begin{minipage}{\linewidth}
		\begin{subfigure}[t]{0.25\textwidth}
			\centering
			\includegraphics[width=3.5cm]{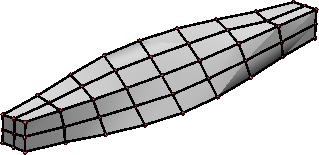}
			\caption{$658$~Hz}
		\end{subfigure}%
		\begin{subfigure}[t]{0.25\textwidth}
			\centering
			\includegraphics[width=3.5cm]{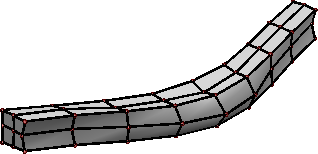}
			\caption{$714$~Hz}
		\end{subfigure}%
		\begin{subfigure}[t]{0.25\textwidth}
			\centering
			\includegraphics[width=3.5cm]{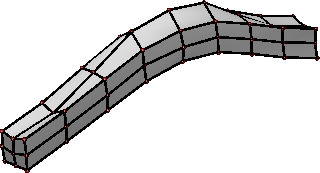}
			\caption{$724$~Hz}
		\end{subfigure}%
		\begin{subfigure}[t]{0.25\textwidth}
			\centering
			\includegraphics[width=3.5cm]{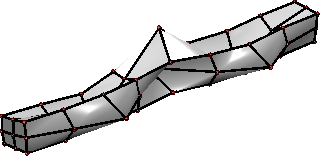}
			\caption{$846$~Hz}
		\end{subfigure}\\
		\begin{subfigure}[t]{0.25\textwidth}
			\centering
			\includegraphics[width=3.5cm]{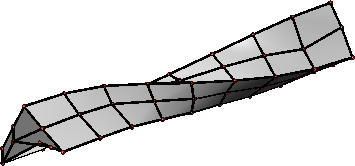}
			\caption{$864$~Hz}
		\end{subfigure}%
		\begin{subfigure}[t]{0.25\textwidth}
			\centering
			\includegraphics[width=3.5cm]{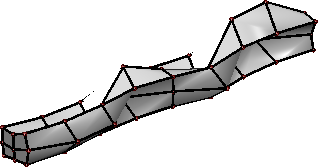}
			\caption{$896$~Hz}
		\end{subfigure}%
		\begin{subfigure}[t]{0.25\textwidth}
			\centering
			\includegraphics[width=3.5cm]{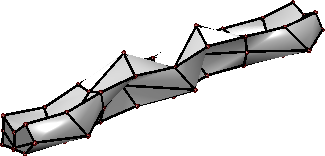}
			\caption{$942$~Hz}
		\end{subfigure}%
		\begin{subfigure}[t]{0.25\textwidth}
			\centering
			\includegraphics[width=3.5cm]{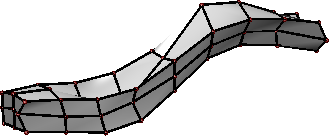}
			\caption{$1102$~Hz}
		\end{subfigure}
	\end{minipage}
	\setcounter{subfigure}{0}
	\begin{minipage}{\linewidth}
		\begin{subfigure}[t]{0.25\textwidth}
			\centering
			\includegraphics[width=3.5cm]{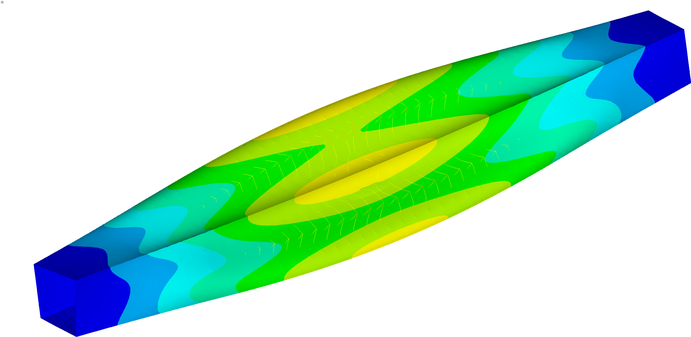}
			\caption{$600.057$~Hz}
		\end{subfigure}%
		\begin{subfigure}[t]{0.25\textwidth}
			\centering
			\includegraphics[width=3.5cm]{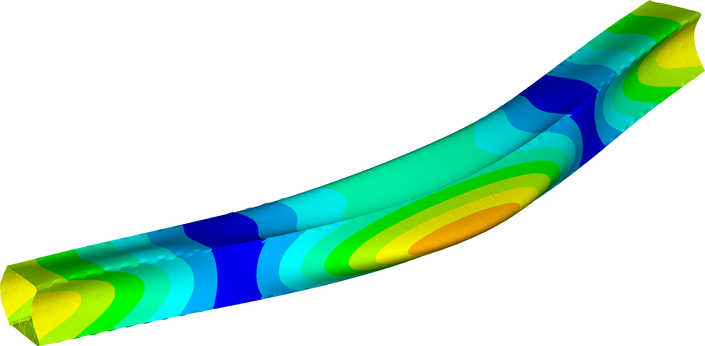}
			\caption{$747.011$~Hz}
		\end{subfigure}%
		\begin{subfigure}[t]{0.25\textwidth}
			\centering
			\includegraphics[width=3.5cm]{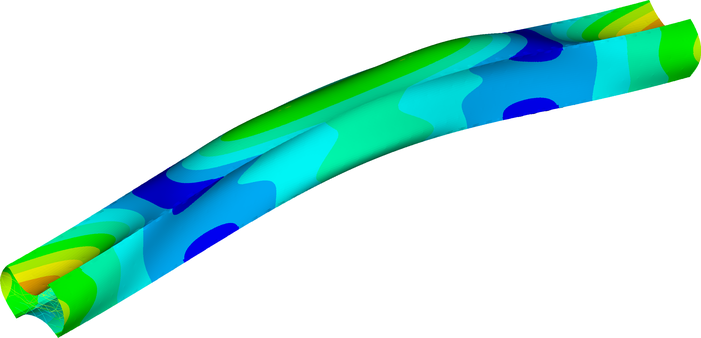}
			\caption{$748.241$~Hz}
		\end{subfigure}%
		\begin{subfigure}[t]{0.25\textwidth}
			\centering
			\includegraphics[width=3.5cm]{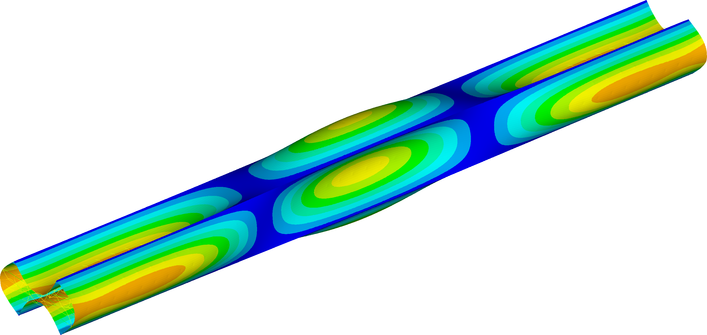}
			\caption{$682.283$~Hz}
		\end{subfigure}\\
		\begin{subfigure}[t]{0.25\textwidth}
			\centering
			\includegraphics[width=3.5cm]{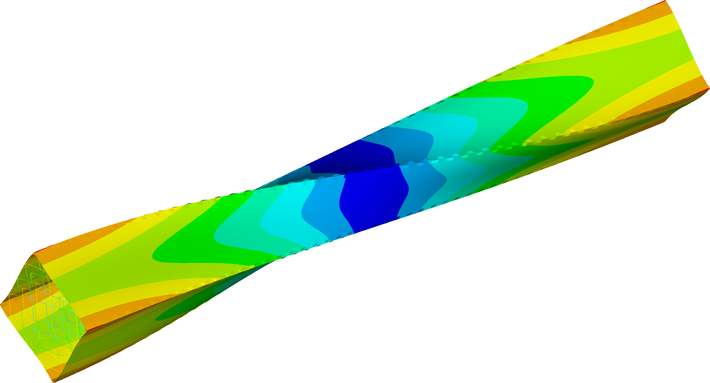}
			\caption{$833.719$~Hz}
		\end{subfigure}%
		\begin{subfigure}[t]{0.25\textwidth}
			\centering	
			\includegraphics[width=3.5cm]{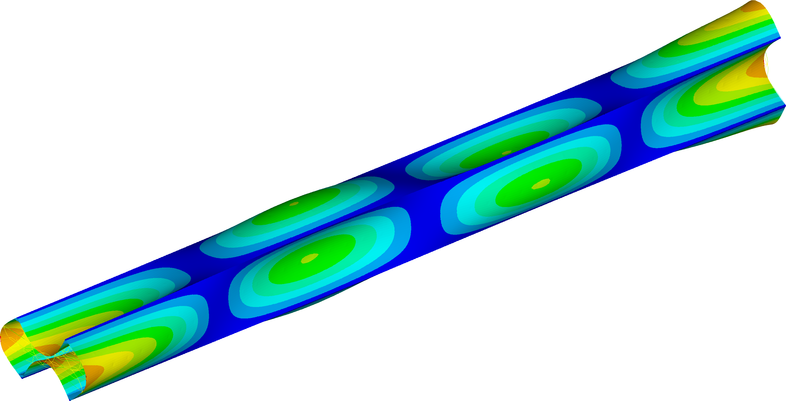}
			\caption{$708.852$~Hz}
		\end{subfigure}%
		\begin{subfigure}[t]{0.25\textwidth}
			\centering
			\includegraphics[width=3.5cm]{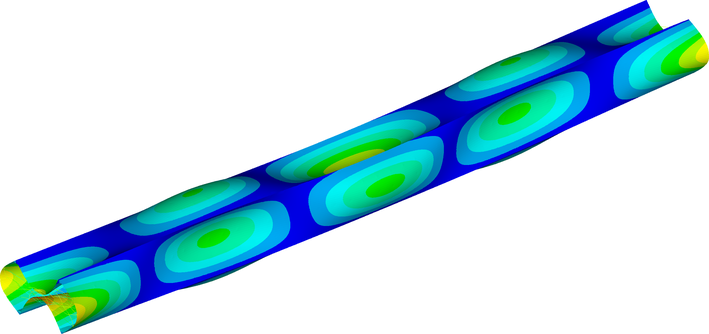}
			\caption{$741.389$~Hz}
		\end{subfigure}%
		\begin{subfigure}[t]{0.25\textwidth}
			\centering
			\includegraphics[width=3.5cm]{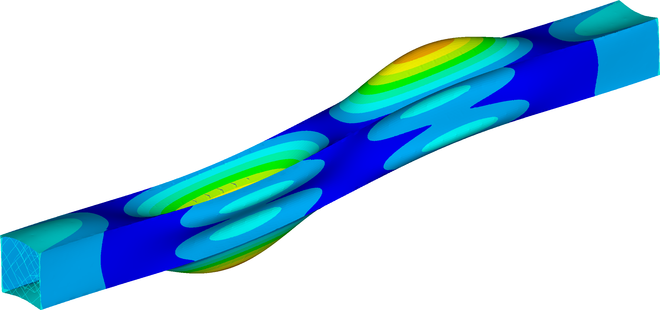}
			\caption{$1019.972$~Hz}
		\end{subfigure}
	\end{minipage}

	\caption{Selected experimentally determined natural frequencies and mode shapes, (a)--(h) top, and finite element model predictions of eigenmodes and eigenfrequencies, (a)--(h) bottom.}
	\label{fig:numdata}\label{fig:expdata}
\end{figure*}

Experimentally determined natural modes and the values of natural eigenfrequencies, Fig.~\ref{fig:expdata} top, were compared with the results of numerical simulations, Fig~\ref{fig:numdata} bottom. Direct comparison in Table~\ref{tab:comparison} reveals\MT{ a}{} sufficient agreement\hl{ of} up to $9 \%$ for eigenfrequencies of shear, bending, and torsional eigenmodes. In the case of buckling, we failed to measure the first and second buckling natural modes, and for the higher eigenmodes\hl{,} the model predictions underestimate the natural frequencies by more than $20 \%$. We attribute these deviations to the overall difficulty of measuring the buckling natural modes and to the manufacturing defects discussed in the previous section.

\begin{table}[!b]
	\centering
	\caption{Comparison of model prediction of eigenfrequencies $f_{\mathrm{FEM}}$ and measured natural frequencies $f_{\mathrm{EXP}}$ using the roving hammer test. Accuracy of individual measurements is denoted by $A$ and the deviation of the model from the experiment by $D$.}
	\label{tab:comparison}
	\scriptsize
	\begin{tabular}{lcccc}
		\hline
		Eigenmode & $f_{\mathrm{FEM}}$ [Hz] & $f_{\mathrm{EXP}}$ [Hz] & $A$ [Hz] & $D$ [$\%$] \\
		\hline
		First shear        & $600.1$  & $658$ & $2$ & $-8.8$ \\
		First bending $y$  & $747.0$  & $714$ & $2$ & $+4.6$ \\
		First bending $z$  & $748.2$  & $724$ & $2$ & $+3.3$ \\
		Third buckling     & $682.3$  & $846$ & $4$ & $-19.3$ \\
		First torsion      & $833.7$  & $864$ & $4$ & $-3.5$ \\
		Fourth buckling    & $708.9$  & $896$ & $4$ & $-20.9$ \\
		Fifth buckling     & $741.4$  & $942$ & $6$ & $-21.3$ \\
		Sixth buckling     & $790.6$  & $1004$ & $6$ & $-21.3$ \\
		Second bending $z$ & $1020.0$ & $1102$ & $6$ & $-7.4$\\
		\hline
	\end{tabular}
\end{table}

\section{Summary and outlook}

This contribution introduces and investigates a unique, fully-automatized \MT{pipeline}{procedure} from an idea to prototyping, with applications to the manufacturing of thin-walled structural composite \MT{tubes}{hollow beams}. In particular, the~considered prototype product is stiffened with a~low weight internal structure designed by an efficient convex linear semidefinite programming formulation. This formulation increased the fundamental free-vibration eigenfrequency above a specified threshold value while avoiding the traditional issue of non-differentiability of multiple eigenvalues~\cite{Achtziger2007}, and limited structural compliance of a compression-molding load case. The optimization output of the non-uniformly distributed lattice-like internal structure was further automatically post-processed and converted into a solid model ready for support-less additive manufacturing.

Our methodology was verified by designing and producing the simply-supported CFRP beam prototype. Optimization yielded an internal structure of $488$~g which increased the fundamental eigenfrequency by $92\%$ and limited the effect of wall instabilities. Moreover, the deflections within the compression-molding load case were limited to $\pm0.5$~mm.

After a successful prototype production, the structural response was validated using the roving hammer test, which showed that bending, torsional, and shear eigenmodes exhibited good agreement with model predictions. For the wall buckling eigenmodes, however, the finite element model underestimated the natural frequencies by almost $22\%$. We attribute this to difficulties in measuring these natural modes and to manufacturing defects associated with compression-molding deformations of the casing.

Improving the structural response with a material more than two orders of magnitude more compliant when compared to CFRP suggests concentrating on substituting ABS with high-stiffness continuous carbon fiber in future studies. Another essential future enhancement resides in accelerating the optimization algorithm by exploiting the range-space sparsity \citep{Kim2011} associated with the segment-based internal-structure decomposition.

\section*{Acknowledgments}

We thank Edita Dvo\v{r}\'{a}kov\'{a} for providing us with her implementation of the \textsc{Mitc4} shell elements~\cite{Dvorakova}, and Ond\v{r}ej Roko\v{s} and Stephanie Krueger for a critical review of the initial versions of this manuscript.

The work of Jan Nov\'{a}k and Robin Poul was supported by the Technology Agency of the Czech Republic, through the project TA\v{C}R TH02020420. Marek Tyburec, Jan Zeman, and Mat\v{e}j Lep\v{s} acknowledge the support of the Czech Science Foundation project No. 19-26143X.

Access to computing and storage facilities owned by parties and projects contributing to the National Grid Infrastructure MetaCentrum provided under the program "Projects of Large Research, Development, and Innovations Infrastructures" (CESNET LM2015042), is greatly appreciated.

\section*{Data availability}
The raw/processed data required to reproduce these findings cannot be shared at this time due to legal or ethical reasons.

\section*{\hl{Appendix A. Static condensation of static LMI}}\label{app:a}
\hl{Consider the equilibrium equation}
\begin{equation}[box=\ybox]
\mathbf{K}(\mathbf{a}) \mathbf{u} = \mathbf{f}\label{eq:system}
\end{equation}
\hl{split into two sets of equations}
\begin{equation}[box=\ybox]
\begin{pmatrix}
\mathbf{K}_a (\mathbf{a}) & \mathbf{K}_b\\
\mathbf{K}_b^\mathrm{T} & \mathbf{K}_c
\end{pmatrix}
\begin{pmatrix}
\mathbf{u}_a\\
\mathbf{u}_b
\end{pmatrix} = 
\begin{pmatrix}
\mathbf{f}_a\\
\mathbf{f}_b
\end{pmatrix},
\end{equation}
\hl{such that only the principal submatrix $\mathbf{K}_a (\mathbf{a})$ depends affinely on $\mathbf{a}$\footnote{\hl{In the context of this article, the matrix $\mathbf{K}_a(\mathbf{a})$ comprises the degrees of freedom of the truss ground structure, $\mathbf{K}_c$ contains the remaining (rotational) degrees of freedom, and $\mathbf{K}_b$ is the coupling term.}}. Assuming that the system~{\eqref{eq:system}} is solvable uniquely for some $\mathbf{a}$, i.e., it holds that $\exists \mathbf{a}\ge\mathbf{0}: \mathbf{K}(\mathbf{a}) \succ \mathbf{0}$, where ``$\succ \mathbf{0}$'' denotes positive definiteness of the left hand side. Note that $\mathbf{a} = \mathbf{1}$ is sufficient for verification that no rigid movement within the structure can occur. Because $\mathbf{K}_c$ is therefore invertible, the degrees of freedom $\mathbf{u}_b$ can be expressed from the second row in terms of $\mathbf{u}_a$}
\begin{equation}[box=\ybox]
\mathbf{u}_b = \left(\mathbf{K}_c\right)^{-1} \mathbf{f}_b - \left(\mathbf{K}_c\right)^{-1} \mathbf{K}_b^\mathrm{T} \mathbf{u}_a \label{eq:ub}
\end{equation}
\hl{and inserted back into the first row,}
\begin{equation}[box=\ybox]
\left[ \mathbf{K}_a (\mathbf{a}) - \mathbf{K}_b \left(\mathbf{K}_c\right)^{-1} \mathbf{K}_b^\mathrm{T}\right] \mathbf{u}_a = \mathbf{f}_a - \mathbf{K}_b \left(\mathbf{K}_c\right)^{-1} \mathbf{f}_b.\label{eq:cond_equilibrium}
\end{equation}
\hl{Structural compliance (work done by external forces) is expressed as}
\begin{equation}[box=\ybox]
c = \mathbf{u}_a^\mathrm{T} \mathbf{f}_a + \mathbf{u}_b^\mathrm{T} \mathbf{f}_b.
\end{equation}
\hl{After inserting {\eqref{eq:ub}} and acknowledging that $\mathbf{K}_c^{-1}$ is Hermitian, we obtain}
\begin{equation}[box=\ybox]
c = 
\mathbf{u}_a^\mathrm{T} 
\left[
\mathbf{f}_a 
- \mathbf{K}_b \left(\mathbf{K}_c\right)^{-1}
 \mathbf{f}_b
\right]
+ \mathbf{f}_b^\mathrm{T} \left(\mathbf{K}_c\right)^{-1} \mathbf{f}_b, \label{eq:cond_comp}
\end{equation}
\hl{i.e., compliance of the condensed problem {\eqref{eq:cond_equilibrium}} and a constant term. Because the compliance of the condensed problem is positive by definition, the constant term represents a~non-negative lower bound on compliances achievable by the internal structure design.}

\hl{Finally, the LMI}
\begin{equation}[box=\ybox]
\begin{pmatrix}
c & -\mathbf{f}^\mathrm{T}\\
-\mathbf{f} & \mathbf{K}(\mathbf{a})
\end{pmatrix} \succeq \mathbf{0}\label{eq:lmi}
\end{equation}
\hl{is equivalent to a smaller LMI}
{\setlength{\mathindent}{0cm}
\begin{equation}[box=\ybox]
\negthickspace\negthickspace\negthickspace%
\begin{pmatrix}
c -\mathbf{f}_b^\mathrm{T} \left(\mathbf{K}_c\right)^{-1} \mathbf{f}_b & 
-\mathbf{f}_a^\mathrm{T} + \mathbf{f}_b^\mathrm{T} \left(\mathbf{K}_c\right)^{-1} \mathbf{K}_b^\mathrm{T}\\
-\mathbf{f}_a + \mathbf{K}_b \left(\mathbf{K}_c\right)^{-1} \mathbf{f}_b & 
\mathbf{K}_a (\mathbf{a}) - \mathbf{K}_b \left(\mathbf{K}_c\right)^{-1} \mathbf{K}_b^\mathrm{T}
\end{pmatrix} \succeq \mathbf{0}.\negthickspace\label{eq:statconlmi}
\end{equation}}%
\hl{Further, if $c > \mathbf{f}_b^\mathrm{T} \left(\mathbf{K}_c\right)^{-1} \mathbf{f}_b$ is a prescribed constant (i.e., not a variable), then {\eqref{eq:statconlmi}} is further reducible, using the Schur complement lemma, e.g., \mbox{\citep[Proposition 16.1]{Gallier2011}}, to a~yet smaller LMI}
\begin{equation}[box=\ybox]
\begin{split}
\mathbf{K}_a (\mathbf{a}) - \mathbf{K}_b \left(\mathbf{K}_c\right)^{-1} \mathbf{K}_b^\mathrm{T} - 
\left(-\mathbf{f}_a^\mathrm{T} + \mathbf{f}_b^\mathrm{T} \left(\mathbf{K}_c\right)^{-1} \mathbf{K}_b^\mathrm{T}\right)\;\;\\
\quad\left(c -\mathbf{f}_b^\mathrm{T} \left(\mathbf{K}_c\right)^{-1} \mathbf{f}_b\right)^{-1}
\left(-\mathbf{f}_a + \mathbf{K}_b \left(\mathbf{K}_c\right)^{-1} \mathbf{f}_b\right) \succeq \mathbf{0}.
\end{split}
\end{equation}

\section*{\hl{Appendix B. Reducing size of free-vibration LMI}}\label{app:b}
\hl{In the case of the free-vibration constraint, we need to directly apply the (generalized) Schur complement lemma. Beginning with reordering of rows and columns, we split the symmetric LMI {\eqref{eq:eigen}} such that only the $\mathbf{K}_a (\mathbf{a})$ and $\mathbf{M}_a (\mathbf{a})$ matrices are functions of $\mathbf{a}$, and the other blocks are constant,}
\begin{equation}[box=\ybox]
\begin{pmatrix}
\mathbf{K}_a (\mathbf{a}) - 4 \pi^2 \overline{f}^2 \mathbf{M}_a(\mathbf{a}) &
\mathbf{K}_b - 4 \pi^2 \overline{f}^2 \mathbf{M}_b \\
\mathbf{K}_b^\mathrm{T} - 4 \pi^2 \overline{f}^2 \mathbf{M}_b^\mathrm{T} &
\mathbf{K}_c - 4 \pi^2 \overline{f}^2 \mathbf{M}_c
\end{pmatrix} \succeq \mathbf{0}.\label{eq:schur_LMI}
\end{equation}
\hl{For the (standard) Schur complement trick we require $\mathbf{K}_c - 4 \pi^2 \overline{f}^2 \mathbf{M}_c \succ \mathbf{0}$ \mbox{\cite[Proposition 16.1]{Gallier2011}}. Since $\mathbf{K}_c \succ \mathbf{0}$ (boundary conditions exclude rigid motions), and $\mathbf{M}_c \succ \mathbf{0}$ by definition, we only need to secure that the fundamental eigenfrequency $f_0$ of the generalized eigenvalue problem}
\begin{equation}[box=\ybox]
 \mathbf{K}_c \mathbf{u}_b - \lambda \mathbf{M}_c \mathbf{u}_b = 0, \label{eq:schur_eig}
\end{equation}
\hl{with $\lambda = 4 \pi^2 f^2$, is strictly greater than $\overline{f}$.}

\hl{Let us therefore first assume that $0 \le \overline{f} < f_0$. Then, the inverse of $\mathbf{K}_c - 4 \pi^2 \overline{f}^2 \mathbf{M}_c$ exists and {\eqref{eq:schur_LMI}} can be rewritten equivalently using the Schur complement lemma into a smaller-sized LMI}
\begin{equation}[box=\ybox]
\begin{split}
\mathbf{K}_a (\mathbf{a}) - 4 \pi^2 \overline{f}^2 \mathbf{M}_a(\mathbf{a}) - \left(\mathbf{K}_b - 4 \pi^2 \overline{f}^2 \mathbf{M}_b\right)\quad\quad\quad \\
\quad\quad\left(\mathbf{K}_c - 4 \pi^2 \overline{f}^2 \mathbf{M}_c\right)^{-1}
\left(\mathbf{K}_b^\mathrm{T} - 4 \pi^2 \overline{f}^2 \mathbf{M}_b^\mathrm{T}\right)
\succeq \mathbf{0}.
\end{split}
\end{equation}

\hl{Second, consider that $f_0 < \overline{f}$. Because the matrix $\mathbf{K}_c - 4 \pi^2 (f_0 + \varepsilon)^2 \mathbf{M}_c$ is indefinite for any $\varepsilon>0$, which renders the original LMI {\eqref{eq:schur_LMI}} infeasible, the eigenfrequency $f_0$ constitutes an upper bound for achievable fundamental eigenfrequencies of the reinforced structure. From the mechanical point of view, the eigenmodes $\mathbf{u}_b$ associated with $f_0$ excite degrees of freedom not reinforced by the internal structure, and therefore the associated eigenfrequencies can not be increased by any admissible internal structure design (given the specific discretization).}

\hl{In the case $\overline{f} = f_0$, reduction of {\eqref{eq:schur_LMI}} relies on the generalized Schur complement lemma \mbox{\cite[Theorem 16.1]{Gallier2011}}, so that {\eqref{eq:schur_LMI}} is equivalent to}
\begin{subequations}
\begin{align}[box=\ybox]
\begin{split}
\mathbf{K}_a (\mathbf{a}) - 4 \pi^2 \overline{f}^2 \mathbf{M}_a(\mathbf{a}) - \left(\mathbf{K}_b - 4 \pi^2 \overline{f}^2 \mathbf{M}_b\right) \qquad\\
\qquad\left(\mathbf{K}_c - 4 \pi^2 \overline{f}^2 \mathbf{M}_c\right)^{\dagger}
\left(\mathbf{K}_b^\mathrm{T} - 4 \pi^2 \overline{f}^2 \mathbf{M}_b^\mathrm{T}\right)
\succeq \mathbf{0},
\end{split}\label{eq:schurG}\\
\begin{split}
\left[\mathbf{I} - \left(\mathbf{K}_c - 4\pi^2 \overline{f}^2 \mathbf{M}_c\right)\left(\mathbf{K}_c - 4\pi^2 \overline{f}^2 \mathbf{M}_c\right)^{\dagger}\right]\qquad\\ 
\qquad\left(\mathbf{K}_b^\mathrm{T} - 4 \pi^2 \overline{f}^2 \mathbf{M}_b^\mathrm{T}\right) = \mathbf{0},\label{eq:schur_gen}
\end{split}
\end{align}
\end{subequations}
\hl{where $\left(\bullet \right)^\dagger$ denotes the Moore-Penrose pseudo-inverse of $\bullet$, and $\mathbf{I}$ is the identity matrix. The second condition {\eqref{eq:schur_gen}} holds iff the columns of $\mathbf{K}_b^\mathrm{T} - 4 \pi^2 \overline{f}^2 \mathbf{M}_b^\mathrm{T}$ are in the image of $\mathbf{K}_c - 4\pi^2 \overline{f}^2 \mathbf{M}_c$. Indeed, {\eqref{eq:schur_gen}} can then be rewritten to}
\begin{equation}[box=\ybox]
\begin{split}
\left[\left(\mathbf{K}_c - 4\pi^2 \overline{f}^2 \mathbf{M}_c\right) - \left(\mathbf{K}_c - 4\pi^2 \overline{f}^2 \mathbf{M}_c\right)\right.\qquad\qquad\\
\quad\quad\left.\left(\mathbf{K}_c - 4\pi^2 \overline{f}^2 \mathbf{M}_c\right)^{\dagger} \left(\mathbf{K}_c - 4\pi^2 \overline{f}^2 \mathbf{M}_c\right) \right] \mathbf{C} = \mathbf{0}.\label{eq:schur_gen2}
\end{split}
\end{equation}
\hl{with the columns of $\mathbf{C}$ being the coefficients of linear combinations of the columns of $\mathbf{K}_c - 4\pi^2 \overline{f}^2 \mathbf{M}_c$, making the term in the square brackets vanish~\mbox{\cite[Lemma 14.1]{Gallier2011}}}.

\hl{Because $\mathrm{Im}( \mathbf{K}_c - 4\pi^2 \overline{f}^2 \mathbf{M}_c ) = \mathrm{Ker}( \mathbf{K}_c - 4\pi^2 \overline{f}^2 \mathbf{M}_c )^\perp$ by \mbox{\cite[Lemma 13.1]{Gallier2011}}, it is spanned by}
\begin{equation}[box=\ybox]
\text{span}\left\{\mathbf{u}_b: \left(\mathbf{K}_c - 4\pi^2 \overline{f}^2 \mathbf{M}_c\right)\mathbf{u}_b = \mathbf{0} \right\}^\perp.
\end{equation}
\hl{Clearly, $\overline{f} = f_0$ might be achieved iff the columns of the coupling term $\mathbf{K}_b^\mathrm{T} - 4 \pi^2 \overline{f}^2 \mathbf{M}_b^\mathrm{T}$ are orthogonal to the eigenmodes occurring in {\eqref{eq:schur_eig}} at $f_0$. From the mechanical point of view, induction of these eigenmodes would result in a~decrease of the associated eigenfrequencies. Note that in practice, equation {\eqref{eq:schur_gen}} can be verified numerically, but it does not guarantee a feasible solution to {\eqref{eq:schurG}}, because other (higher) eigenfrequencies associated with eigenmodes of {\eqref{eq:schur_eig}} may decrease below $f_0$ due to the coupling term.}

\bibliography{article}
\bibliographystyle{elsarticle-num-names}

\end{document}